\documentclass[aps,pre,twocolumn,superscriptaddress,showpacs]{revtex4-1}

\usepackage{amsmath}
\usepackage{amssymb}
\usepackage{hyperref}
\usepackage[arrow, matrix, curve]{xy}
\usepackage{bm}
\usepackage{yhmath}
\usepackage{color}
\usepackage{graphicx}
\newcommand\diff{\mathrm{d}}

\DeclareMathOperator\Imag{Im}

\usepackage{url,hyperref}
\usepackage[usenames,dvipsnames]{xcolor}
\hypersetup{colorlinks=true, linkcolor=BrickRed, urlcolor=blue!50!black, citecolor=blue!50!black}

\usepackage[normalem]{ulem}

\newcommand{\norm}[1]{\left\lVert #1 \right\rVert}

\begin{document}


\title{Tagged-particle motion in a dense confined liquid}




\author{Simon Lang}
\affiliation{Institut f\"ur Theoretische Physik, Leopold-Franzens-Universit\"at Innsbruck, Technikerstra{\ss}e~25/2,  A-6020 Innsbruck, Austria}
\affiliation{Institut f\"ur Theoretische Physik, Friedrich-Alexander-Universit\"at Erlangen-N\"urnberg, Staudtstra{\ss}e~7, 91058, Erlangen, Germany}


\author{Thomas Franosch}
\affiliation{Institut f\"ur Theoretische Physik, Leopold-Franzens-Universit\"at Innsbruck, Technikerstra{\ss}e~25/2,  A-6020 Innsbruck, Austria}
\affiliation{Institut f\"ur Theoretische Physik, Friedrich-Alexander-Universit\"at Erlangen-N\"urnberg, Staudtstra{\ss}e~7, 91058, Erlangen, Germany}




\date{\today}

\begin{abstract}
We investigate the dynamics of a tagged particle embedded in a strongly interacting confined liquid enclosed between two opposing flat walls. Using the Zwanzig-Mori projection operator formalism we obtain an equation of motion for the incoherent scattering function  suitably generalized to account for the lack of translational symmetry. We close the equations of motion by a self-consistent  mode-coupling ansatz. The   interaction of the tracer with the surrounding liquid is encoded in generalized direct correlation functions.  We extract the in-plane dynamics and provide a microscopic expression for the diffusion coefficient parallel to the walls. The solute particle may differ in size or interaction from the surrounding host-liquid constituents offering the possibility of a systematic analysis of dynamic effects on the tagged-particle motion in confinement.
\end{abstract}


\maketitle


\section{INTRODUCTION}
The dynamics of dense liquids is dominated by  the cage effect, where a particle is transiently trapped by its surrounding neighbors.
 The rattling in the cages and the escape thereof induces strong memory effects at microscopic time and length scales~\cite{Goetze:Complex_Dynamics}.
In very dense liquids the interactions between the particles are so strong, that the liquid may finally end up in an  arrested state at macroscopic time scales, which is referred to as a glass.
 In this dense regime close to the glass transition,  transport is highly correlated and the dynamics slows down  dramatically by friction, in particular, the self-diffusion becomes intriguingly small and the viscosity increases by orders of magnitude.

A self-consistent ansatz for the intermediate scattering functions is provided by the mode-coupling theory (MCT) of the glass transition~\cite{Bengtzelius:1984,Goetze:Complex_Dynamics} and characteristic features of the slowing down of motion  due to the cage dynamics are rationalized.
In the vicinity of the  glass transition singularity  a universal scenario is predicted by MCT, which is corroborated by numerous experiments and simulations~\cite{Li:1992,Megen:1993a,Wuttke:1994,Torre:2000,Megen:1998,Singh:1998,Goetze:1999,Goetze:Complex_Dynamics,Kob:1994,Kob:1995a,Kob:1995b,Kaemmerer:1998,Kaemmerer:1998a}.
In particular,  the regime of the $\beta$ relaxation close to the plateau values with two time fractals equipped with nontrivial exponents are uncovered as asymptotic laws of the underlying equations~\cite{Goetze:Complex_Dynamics,Franosch:1997,Fuchs:1998}.

Novel phenomena occur if dense liquids are exposed to confinement on the molecular scale. Then, the new length scale competes with the typical interaction range of the particles and imposes local packing frustrations, which affect the structure and dynamics of the liquid. For instance, the dynamics of liquids within a frozen disorder~\cite{Krakoviack:2005,*Krakoviack:2007,*Krakoviack:2011,Kurzidim:2009,Kim:2009,Horbach:2002,Voigtmann:2006,Szamel:2013} has been investigated and an intriguing interplay of localization phenomena and the glass transition of the liquid has been unraveled.

In contrast, external walls can be utilized to squeeze dense liquids into narrow pores, channels or wedges.
A multitude of simulations and experiments address the question of how confinement affects the dynamics of dense liquids  for various  particle-wall interactions or roughness of the walls~\cite{Fehr:1995,Scheidler:2000b,*Scheidler:2000a,*Scheidler:2002,*Scheidler:2004,Teboul:2002,Varnik:2002,Baschnagel:2005,Nugent:2007,Eral:2009,Eral:2011,Edmond:2012,Gallo:2000a,*Gallo:2000b,*Gallo:2012,Krishnan:2012}. Relaxation times or transport coefficients such as the diffusion coefficient are then suitable measures to characterize the rapidity of particle motion.
In particular, for a dense hard-sphere liquid the diffusivity has been found to oscillate as a function of the slit width, which is a dynamical manifestation of (in)commensurability effects of packing in strong confinement~\cite{Mittal:2008}. Beyond recent successful investigations to link static thermodynamic observables directly to transport coefficients~\cite{Mittal:2006,*Mittal:2007a,*Mittal:2007b,Goel:2008,Goel:2009,Krekelberg:2013,Ingebrigtsen:2013}, an explanation for this phenomenon in the dense regime by a microscopic theory of dynamical origin is yet lacking.

Autocorrelation functions of the phase-space coordinates of a single distinguished particle  are usually referred to as self-correlation functions and the dynamics  as self-dynamics. The fundamental statistical quantity encoding the self-dynamics of a tagged particle is the incoherent scattering function, which can be measured by several techniques including neutron scattering, dynamic light scattering, single-particle-tracking experiments or, in particular, by computer simulations~\cite{Kob:1994,*Kob:1995a,*Kob:1995b,Kaemmerer:1998,Kaemmerer:1998a,Voigtmann:2004}. Provided the incoherent scattering function is known, all moments of the displacement can be extracted, in particular, the mean-square displacement providing the diffusion coefficient for sufficiently long times.

For a tagged particle meandering in a  densely packed liquid a self-consistent description for the incoherent scattering function is also provided by the MCT~\cite{Fuchs:1998}.  The MCT functional, which generates the slow critical dynamics due to the formation of local cages, is a linear functional of the incoherent scattering function of the tracer and the collective scattering function of the host liquid. The interactions with the surrounding solvent are provided by static vertices and  explicitly encoded in a direct correlation function. The coherent dynamics merely enters as an \emph{a priori} known input quantity and, in principle, no restriction on its origin is imposed.
The tagged-particle dynamics in the framework of MCT has been extensively elaborated ranging from simple bulk systems~\cite{Fuchs:1998,Voigtmann:2004}, binary mixtures~\cite{Flenner:2005,Voigtmann:2011,Hajnal:2011}, sheared liquids~\cite{Krueger:2011,Harrer:2012}, granular fluids~\cite{Sperl:2012}, molecules~\cite{Franosch:1997,Chong:2001}, polymeric liquids~\cite{Chong:2002}, and tagged-particle motion in porous disorder~\cite{Krakoviack:2009,Spanner:2013}.
Extending the theory to inhomogeneous situations has been done in the case of the collective dynamics between two flat walls~\cite{Lang:2010,Lang:2012}, for local external fields ~\cite{Biroli:2006} and periodic potential landscapes~\cite{Nandi:2011}.

In the present work we derive equations of motion for the incoherent scattering function of a tagged particle of arbitrary size in a densely packed confined liquid. The incoherent scattering function is suitably generalized to account for the lack of translational invariance perpendicular to the channel. We exploit the projection-operator formalism to obtain formally exact Zwanzig-Mori equations of motion and close the formalism by a mode-coupling ansatz. We employ  a suitably designed Ornstein-Zernike equation to quantify the coupling strength to the host liquid via a direct correlation function. To establish contact to recent findings on the oscillatory behavior of the diffusion coefficient as a function of the plate distances~\cite{Mittal:2008} we extract the in-plane dynamics and derive a microscopic expression for the diffusion coefficient along the channel.

\section{MODEL}
\begin{figure}
 \centering
  \raisebox{-0.5\height}{\includegraphics[width=0.2\linewidth]{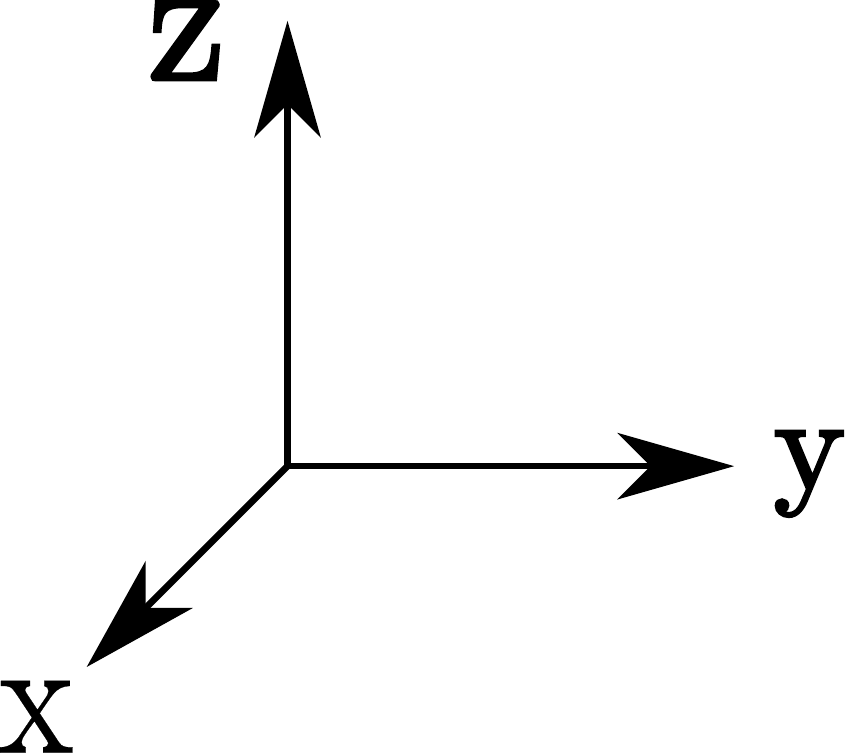}}
  \raisebox{-0.5\height}{\includegraphics[width=0.75\linewidth]{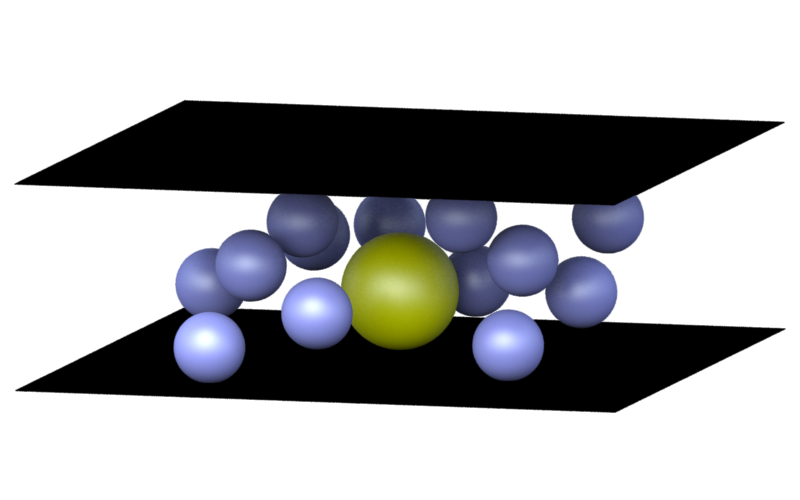}}
\caption{Tagged particle (yellow) surrounded by a dense solvent (blue particles) in a planar slit pore.}
\label{fig:tagged}
\end{figure}
We consider a liquid of $N$ structureless identical solvent particles of mass $m$ confined to a slit pore (see Fig.~\ref{fig:tagged}), and add an additional particle of mass $m_{s}$ to the system. This particle shall be in the following referred to as the tagged particle or the tracer.  The accessible volume for the tracer is given by $V_{s}=AL_{s}$, where $A$ refers to the lateral area and $L_{s}$ corresponds to the effective transversal distance for the tagged particle between the walls. The tracer particle may differ in size and interaction from the remaining solvent particles; in particular the effective confinement length in perpendicular direction is not required to coincide with the accessible transversal distance $L$ for the host-liquid particles, i.e., $L_{s}\neq L$.  The total Hamilton function consists of a mere  solvent contribution $H(\{\vec{x}_{n}\},\{\vec{p}_{n}\})$ with the coordinates of the solvent particles $\{\vec{x}_{n}= (\vec{r}_{n},z_{n})\}$ and a contribution arising from the tracer $H_{s}(\vec{x}_{s},\vec{p}_{s},\{\vec{x}_{n}\})$,  coordinates $\vec{x}_{s}= (\vec{r}_{s},z_{s})$,
\begin{equation}
H_{\text{tot.}}=H(\{\vec{x}_{n}\},\{\vec{p}_{n}\})+H_{s}(\vec{x}_{s},\vec{p}_{s},\{\vec{x}_{n}\}).
\end{equation}
Here and in the following all quantities characterizing the self-dynamics are indicated by the index $s$. The solvent Hamiltonian follows from Ref.~\cite{Lang:2012} to
\begin{equation}
 H(\{\vec{x}_{n}\},\{\vec{p}_{n}\})=\sum_{n=1}^{N} \frac{\vec{p}_{n}^2}{2m}+V(\{\vec{x}_{n}\})+U(\{z_{n}\}),
\end{equation}
where mutual pairwise interactions between the solvent particles are assumed
\begin{equation}
V(\{\vec{x}_{n}\})=\sum_{n<m}^{N} \mathcal{V}(|\vec{x}_{n}-\vec{x}_{m}|).
\end{equation}
The interactions with the walls are described by single-particle potentials
\begin{equation}
U(\{z_{n}\})=\sum_{n=1}^{N}\mathcal{U}(z_{n}),
\end{equation}
where
\begin{equation}
\mathcal{U}(z)=
\begin{cases}
    \mathcal{U}^{W}(z) & \text{for } |z| \leq L/2,  \\
  \infty & \text{for } |z| > L/2.
\end{cases}
\end{equation}
In the case of identical walls, inversion symmetry $\mathcal{U}^W(-z) = \mathcal{U}^W(z)$ is satisfied. Similarly, an additional contribution from the tracer particle enters
\begin{align}
 H_{s}(\vec{x}_{s},\vec{p}_{s},\{\vec{x}_{n}\})=\frac{\vec{p}_{s}^2}{2m_{s}}+\sum_{n=1}^{N} \mathcal{V}_{s}(|\vec{x}_{n}-\vec{x}_{s}|)+\mathcal{U}_{s}(z_{s}).
\end{align}
In particular, the interaction of the tagged particle with the liquid particles solely depends on their mutual distance. The tagged particle interacts with the walls via
\begin{equation}
\mathcal{U}_{s}(z)=
\begin{cases}
    \mathcal{U}^{W}_{s}(z) & \text{for } |z| \leq L_{s}/2,  \\
  \infty & \text{for } |z| > L_{s}/2,
\end{cases}
\end{equation}
where similar to the host liquid~\cite{Lang:2012} additional specific tracer-wall interactions such as adsorption induced by hydrophilic or adhesive, respectively, hydrophobic or cohesive forces, are thereby included.

The time evolution of the tagged particle shall be determined by Newton's equation of motion. A local variable specifying its position in space is the microscopic density,
\begin{equation}
\rho^{(s)}(\vec{r},z,t)=\delta[\vec{r}-\vec{r}_{s}(t)]\delta[z-z_{s}(t)].
\end{equation}
Averaged  observables reflect the symmetries of the confining space; see Ref.~\cite{Lang:2012} for the case of two opposing flat hard walls. Here, for instance, the average tagged-particle density,
\begin{equation}\label{eq:tdens}
 \langle \rho^{(s)}(\vec{r},z,t) \rangle=:N^{-1} n^{(s)}(z),
\end{equation}
depends due to translational invariance along the container walls only on the $z$ coordinate similar to the density profile of the solvent: $n(z)=\sum_{n=1}^{N}\langle \delta[\vec{r}-\vec{r}_{n}(t)]\delta[z-z_{n}(t)] \rangle$~\cite{Lang:2012}. Thermal averages $\langle \cdot \rangle$  are performed in the canonical ensemble associated with the phase space spanned by the $N+1$ particles.  Assuming the tagged particle to be of the same species as the solvent component, the equivalence  $n^{(s)}(z)\equiv n(z)$ follows in the thermodynamic limit: $N,A \to \infty$ at fixed area density $n_0=N/A$. The density profile of the tracer $n^{(s)}(z)$ can then be written as $n^{(s)}(z)=n_{0}\langle\delta(z-z_{s}) \rangle$.

Suitable eigenmodes for the planar geometry are provided by the plane wave basis $\exp [-i Q_{\mu}^{(s)}z]$ with $Q_{\mu}^{(s)}=2\pi \mu/L_{s}, \mu \in \mathbb{Z}$, which are complete $\sum_{\mu}\exp [i Q_{\mu}^{(s)}z] \exp [-i Q_{\mu}^{(s)}z']=L_{s}\delta(z-z')$ and orthogonal $L^{-1}_{s}\int_{-L_{s}/2}^{L_{s}/2} \diff z\ \exp [i Q_{\mu}^{(s)} z] \exp [-i Q_{\nu}^{(s)} z]=\delta_{\mu\nu}$.
The Fourier components of the tagged particle`s density profile can be obtained by
\begin{equation}
 n^{(s)}_{\mu}= \int_{-L_{s}/2}^{L_{s}/2} \diff z\ n^{(s)}(z) e^{i Q_{\mu}^{(s)} z}.
\end{equation}
 Since the profile $n^{(s)}(z)$ is real, $n^{(s)}_{\mu}=n^{(s)*}_{-\mu}$, and for symmetric walls    $n^{(s)}_{\mu}=n^{(s)}_{-\mu}$ holds as well. Furthermore, the zero-element $n^{(s)}_{0}$ of the density modes always coincides with the planar density $n^{(s)}_{0}=n_{0}$ of the solvent, where a correction of order $1/N$ has been neglected. 

The tagged-particle current density can be decomposed into a component parallel and perpendicular to the walls. The parallel part specifies the in-plane momentum of the particle
\begin{equation}
 \vec{j}^{\parallel, (s)}(\vec{r},z,t)= \vec{v}_{s}(t) \delta [\vec{r}-\vec{r}_{s}(t)] \delta [z-z_{s}(t)],
\end{equation}
and the perpendicular part the momentum in transversal direction
\begin{equation}
 j^{\perp ,(s)}(\vec{r},z,t)= v^{z}_{s}(t) \delta [\vec{r}-\vec{r}_{s}(t)] \delta [z-z_{s}(t)],
\end{equation}
where the components of the tracer particle velocity $\vec{p}_{s}(t)=m_{s}(\vec{v}_{s}(t),v^{z}_{s}(t))$ has been introduced.
Since the probability to find the tracer particle in the slit is conserved, a continuity equation can be formulated
\begin{equation}
 \partial_{t} \rho^{(s)}(\vec{r},z,t)+\vec{\nabla}_{\vec{r}} \cdot \vec{j}^{\parallel,(s)}(\vec{r},z,t)+\nabla_{z}j^{\perp,(s)}(\vec{r},z,t)=0,
\end{equation}
with $\vec{\nabla}=(\vec{\nabla}_{\vec{r}},\nabla_{z})$.

The spatial dependence in lateral direction is decomposed into ordinary plane waves, $\text{e}^{\text{i} \vec{q} \cdot \vec{r}}$, where the wave vectors $\vec{q} = (q_x,q_y)$ are treated
initially as discrete $ (q_x,q_y) \in (2\pi/\sqrt{A}) \mathbb{Z}^2$. As usual, the thermodynamic limit then allows us to replace $(1/A) \sum_{\vec{q}} \ldots \to  (2\pi)^{-2}\int \diff^2 \vec{q}  ...$. In transverse direction, the decomposition is rendered with the aid of the commensurate discrete Fourier modes. The Fourier components of the microscopic tagged-particle density are then given by
\begin{equation}
\rho_\mu^{(s)}(\vec{q},t)= e^{\text{i}\vec{q}\cdot \vec{r}_{s}(t)} \exp [\text{i} Q_{\mu}^{(s)} z_{s}(t)].
\end{equation}
Similarly, the longitudinal current density of the tagged particle is decomposed into the plane-wave basis
\begin{equation}
 j^{\alpha,(s)}_{\mu}(\vec{q},t)= b^{\alpha}(\vec{v}_{s}(t)\cdot \hat{\vec{q}},v_{s}^{z}(t)) e^{\text{i}\vec{q}\cdot \vec{r}_{s}(t)} \exp [i Q_{\mu}^{(s)} z_{s}(t)],
\end{equation}
where we employ  the selector $b^\alpha(x,z) = x \delta_{\alpha,\parallel} + z\delta_{\alpha,\perp}$ and indicate the unit vector by $\hat{\vec{x}}= \vec{x}/x$ for a compact notation. Then, the continuity equation can be recast to
\begin{equation}\label{eq:continuity}
 \partial_t \rho_\mu^{(s)} (\vec{q},t)= \text{i} \sum_{\alpha=\parallel,\perp} b^{\alpha}(q,Q_{\mu}^{(s)}) j_{\mu}^{\alpha,(s)}(\vec{q},t).
\end{equation}

The relevant correlation functions in the present context are built by these phase-space functions. In particular, the incoherent scattering function readily follows to
\begin{equation}
S_{\mu \nu}^{(s)}(q,t)= \langle \delta \rho_{\mu}^{(s)}(\vec{q},t)^{*} \delta \rho_{\nu}^{(s)}(\vec{q})\rangle,
\end{equation}
which encodes the dynamics of the tagged particle in the confined geometry containing all moments of displacement. Here, the convention $\delta \rho_{\mu}^{(s)}(\vec{q},t):=  \rho_{\mu}^{(s)}(\vec{q},t)-\langle \rho_{\mu}^{(s)}(\vec{q},t)\rangle$ is adopted and we have abbreviated the initial density mode by $ \rho_{\mu}^{(s)}(\vec{q})\equiv\rho_{\mu}^{(s)}(\vec{q},t=0)$. From translational symmetry lateral to the walls one infers $\langle \rho_{\mu}^{(s)}(\vec{q},t)\rangle=\delta_{\vec{q},0} n_{\mu}^{(s)}/n_{0}$, which is non zero for $\vec{q}\equiv 0$, only. The incoherent scattering function is suitably generalized to account for the lack of translational symmetry perpendicular to the walls.
From the space-time symmetries analyzed in Ref.~\cite{Lang:2012}, one infers that the incoherent scattering function only depends on the modulus  $q=|\vec{q}|$. Its real-space representation is the density-density auto-correlation function $G^{(s)}(|\vec{r}-\vec{r}'|,z,z',t)=A\langle \delta \rho^{(s)}(\vec{r},z,t) \delta \rho^{(s)}(\vec{r}',z') \rangle$ and both are connected via the subsequent Fourier transform,
\begin{align} \label{eq:Fourierdef}
 S_{\mu\nu}^{(s)}(q,t)=&
 \int\limits_{-L_{s}/2}^{L_{s}/2}\!\!\!\!\diff z \!\! \int\limits_{-L_{s}/2} ^{L_{s}/2}\!\!\!\!\diff z^{\prime}\int\limits_{A} \! \diff (\vec{r}-\vec{r}') G^{(s)}({|\vec{r}-\vec{r}^{\prime}|,z,z^{\prime},t})\nonumber  \\
 & \times \exp\left[- \text{i}(Q_\mu^{(s)} z- Q^{(s)}_\nu z') \right] \text{e}^{-\text{i} \vec{q} \cdot (\vec{r}-\vec{r}^{\prime})}.
\end{align}
 Further relations for the matrix elements can be elaborated by  utilizing time-inversion and time-translational symmetry, in particular the incoherent scattering function is Hermitian,
\begin{equation}
 S_{\mu\nu}^{(s)}(q,t) =S_{\mu\nu}^{(s)}(q,-t) = S_{\nu\mu}^{(s)}(q,t)^*.
\end{equation}
For symmetric walls, the  inversion symmetry additionally yields
\begin{equation}
 S_{\mu\nu}^{(s)}(q,t) =S_{\mu\nu}^{(s)}(q,t)^* = S_{-\mu-\nu}^{(s)}(q,t),
\end{equation}
i.e., the matrices are real symmetric and invariant under simultaneous change of the sign of the mode index. For $\vec{q}\neq 0$ the initial value of the incoherent scattering function $S_{\mu \nu}^{(s)}(q,t=0)=:S_{\mu \nu}^{(s)}$ is entirely determined by the density modes of the tagged particle,
\begin{equation}
S_{\mu \nu}^{(s)}=n^{(s)*}_{\mu-\nu}/n_{0}.
\end{equation}
The dependence on the discrete Fourier modes $\mu,\nu$ is a manifestation of the breaking of translational symmetry perpendicular to the walls and stands in strong contrast to simple bulk liquids~\cite{Fuchs:1998} or systems with disorder average~\cite{Krakoviack:2009}. 
The correlations of the tagged-particle currents involve matrices with channel indices $\alpha,\beta \in \{\parallel,\perp\}$ referring to the currents parallel and perpendicular to the walls,
\begin{equation}\label{eq:curr}
\mathcal{J}_{\mu \nu }^{\alpha \beta,(s)}(\vec{q},t)= \langle  j_{\mu}^{\alpha,(s) }(\vec{q},t)^* j_{\nu}^{\beta,(s)}(\vec{q})\rangle.
\end{equation}
Symmetries of the tagged-particle current correlator can be taken over from Ref.~\cite{Lang:2012}, in particular, one infers
\begin{equation}
\mathcal{J}^{\alpha\beta,(s)}_{\mu\nu}(q,t) = \mathcal{J}^{\alpha\beta,(s)}_{\mu\nu}(q,-t) = \mathcal{J}^{\beta\alpha,(s)}_{\nu\mu}(q,t)^*,
\end{equation}
and for symmetric walls, additionally,
\begin{equation}
 \mathcal{J}^{\alpha\beta,(s)}_{\mu\nu}(q,t) = \mathcal{J}^{\alpha\beta,(s)}_{\mu\nu}(q,t)^* = \mathcal{J}^{\beta\alpha,(s)}_{-\nu-\mu}(q,t).
\end{equation}
A decoupling property follows in the long-wavelength limit $q\to 0$ for the tagged-particle current-correlator  $\lim_{q\to 0} \mathcal{J}^{\alpha\beta,(s)}_{\mu\nu}(q,t)=:\delta^{\alpha \beta}\mathcal{J}^{\alpha,(s)}_{\mu\nu}(t)$ due to the symmetry under rotations around the $z$ axis.
To initial time $t=0$ the tagged-particle current correlator  decouples as well with respect to the $\alpha,\beta$ indices, $\mathcal{J}^{\alpha\beta,(s)}_{\mu\nu}(q,t=0)=: \mathcal{J}^{\alpha,(s)}_{\mu\nu} \delta^{\alpha \beta}$, and one finds explicitly,
\begin{equation}
\mathcal{J}^{\alpha,(s)}_{\mu\nu}= (v^{(s)}_{\text{th}})^2 S_{\mu \nu}^{(s)},
\end{equation}
where the thermal velocity of the tagged particle $v^{(s)}_{\text{th}}=(k_{B}T/m_{s})^{1/2}$ is introduced.

\section{THEORY FOR THE INCOHERENT SCATTERING FUNCTION (MCT)}
\subsection{Zwanzig-Mori equations of motion}
In this subsection we derive equations of motion for the incoherent scattering function following the Zwanzig-Mori technique. Applying this strategy similar equations of motion as in the case of the coherent dynamics in confined geometry are obtained~\cite{Lang:2012}. In the following we present this derivation to provide a self-contained description.

The time evolution of the tagged-particle dynamics shall obey Newton's equation of motion, which implies that phase-space functions $A(t) \equiv A(\{\vec{p}_{j}(t)\}, \{\vec{x}_{j}(t)\})$ with $j=1,2,\dots,N+1$ are driven by the Liouville operator ${\cal L}$,  $\partial_t A(t) = \{ A(t), H \} \equiv \text{i} {\cal L} A(t)$~\cite{Hansen:Theory_of_Simple_Liquids}.  A formal solution of this equation is given by $A(t)  = \exp(\text{i} {\cal L} t) A$, where  we
suppress  the argument  if the phase-space function is evaluated at initial time $t=0$, so $A=A(t=0)$.

The Hilbert space of the set of fluctuating phase space functions possesses an inner product $\langle  A |  B \rangle \equiv \langle \delta A^{*} \delta B\rangle$ as correlation functions between fluctuations  $\delta A = A - \langle A \rangle$~\cite{Goetze:Complex_Dynamics}. By the hermiticity of the Liouville operator with respect to this inner product, time-correlation functions are represented by matrix elements $\langle \delta A(t)^* \delta B \rangle = \langle A | {\cal R}(t) | B \rangle$ of the backwards-time evolution operator ${\cal R}(t) = \exp(- \text{i} {\cal L} t)$.  Differentiating  the backwards-time evolution operator with respect to $t$ yields an equation of motion $\partial_t \mathcal{R}(t) = -\text{i} \mathcal{L} \mathcal{R}(t)$.

The Hilbert space of the phase-space functions shall be partitioned by employing projection operators ${\cal P}$ selecting distinguished vectors of this space. Thereby completeness is assured by the orthogonal complement  ${\cal Q}$ satisfying the identity ${\cal P}+{\cal Q}=\bm{1}$.
A formally exact equation of motion for the backwards-time evolution operator is readily obtained; see, for instance, Appendix B in Ref.~\cite{Lang:2012},
\begin{align}\label{eq:master}
&\partial_t\mathcal{P} \mathcal{R}(t)\mathcal{P}+\text{i}\mathcal{P} \mathcal{L}\mathcal{P} \mathcal{R}(t)\mathcal{P} \\ \nonumber
&+\int_{0}^{t} \diff t' \mathcal{P}\mathcal{L}\mathcal{Q}  R'(t-t') {\cal Q } {\cal L}\mathcal{P}\mathcal{R}(t')\mathcal{P}=0,
\end{align}
which is the starting point of the projection-operator formalism. Here, $R'(t)=e^{-i \mathcal{Q} \mathcal{L} \mathcal{Q} t}$ is referred to as the reduced backwards time-evolution operator, which drives the dynamics via the reduced Liouville operator  $\mathcal{L}'=\mathcal{Q}\mathcal{L}\mathcal{Q}$  of the subspace spanned by the orthogonal complement ${\cal Q}=\bm{1}-{\cal P}$. Sandwiching Eq.~\eqref{eq:master} with the tagged-particle density  $| \rho_{\nu}^{(s)}(\vec{q})\rangle$ as distinguished variable and choosing the projector onto the same subspace $\mathcal{P}_{\rho^{(s)}}=\sum_{\kappa \lambda}|\rho_{\kappa}^{(s)}(\vec{q})\rangle [(\mathbf{S}^{(s)})^{-1}]_{\kappa \lambda}\langle\rho_{\lambda}^{(s)}(\vec{q})|=\bm{1}-\mathcal{Q}_{\rho^{(s)}}$  one obtains an integro-differential equation for the incoherent scattering function  $S^{(s)}_{\mu \nu}(q,t)=\langle \rho^{(s)}_{\mu}(\vec{q})|\mathcal{R}(t)|\rho_{\nu}^{(s)}(\vec{q})\rangle$, which is of Volterra type
\begin{equation}\label{eq:eom1}
 \dot{S}_{\mu\nu}^{(s)}(q,t) + \sum_{\kappa\lambda}\!\! \int_0^t  \! \! K_{\mu \kappa}^{(s)}(q,t-t')[(\mathbf{S}^{(s)})^{-1}]_{\kappa \lambda} S_{\lambda\nu}^{(s)}(q,t')\diff t' = 0.
\end{equation}
The term $\langle \rho^{(s)}_{\mu}(\vec{q})|\mathcal{P} \mathcal{L}\mathcal{P} \mathcal{R}(t)\mathcal{P}| \rho^{(s)}_{\nu}(\vec{q}) \rangle\equiv 0$ vanishes due to time-inversion symmetry. Generally, the equation is also called the Zwanzig-Mori equation of motion and is  naturally equipped by a memory function, which accounts for the entire history of the dynamical process. Explicitly, the kernel is given by
\begin{equation}\label{eq:K}
 K_{\mu\nu}^{(s)}(q,t)= \langle \mathcal{L} \rho_{\mu}^{(s)}(\vec{q}) | \text{e}^{-\text{i}\mathcal{Q}_{\rho^{(s)}} \mathcal{L} \mathcal{Q}_{\rho^{(s)}} t }  |
\mathcal{L} \rho^{(s)}_{\nu}(\vec{q})\rangle,
\end{equation}
and by utilizing the continuity equation [see Eq.~\eqref{eq:continuity}], it splits due to the presence of two tagged-particle currents,
\begin{equation}\label{eq:contract_t}
K_{\mu\nu}^{(s)}(q,t)= \sum_{\alpha\beta=\parallel,\perp}b^{\alpha}(q,Q_{\mu}^{(s)}) \mathcal{K}^{\alpha\beta,(s)}_{\mu\nu}(q,t)b^{\beta}(q,Q_{\nu}^{(s)}).
\end{equation}
The kernel $\mathcal{K}^{\alpha\beta,(s)}_{\mu\nu}(q,t)$ is constructed by the current observables $|j^{\beta,(s)}_{\nu}(\vec{q})\rangle$ and is specified as the reduced tagged-particle current correlator,
\begin{equation}
\mathcal{K}^{\alpha \beta,(s)}_{\mu \nu}(q,t)= \langle j_{\mu}^{\alpha,(s)}(\vec{q}) |\text{e}^{-\text{i}\mathcal{Q}_{\rho^{(s)}} \mathcal{L} \mathcal{Q}_{\rho^{(s)}} t }  |j_{\nu}^{\beta,(s)}(\vec{q})\rangle.
\end{equation}
The symmetries are literally the same as found for $\mathcal{J}^{\alpha \beta,(s)}_{\mu\nu}(q,t)$, see Eq.~\eqref{eq:curr}. In the limit of small wave numbers $q\to 0$ it satisfies the decoupling property with respect to the channel indices $\lim_{q\to 0} \mathcal{K}^{\alpha\beta,(s)}_{\mu\nu}(q,t)=: \delta^{\alpha\beta}\mathcal{K}^{\alpha,(s)}_{\mu\nu}(t)$. For the particular indices $\lim_{q\to 0} \mathcal{K}^{\parallel\parallel,(s)}_{00}(q,t)=Z_{\parallel}(t)$ one recovers the velocity-velocity auto-correlation function parallel to the walls
\begin{equation}\label{eq:VACF}
Z_{\parallel}(t)=\langle \hat{\vec{q}}\cdot \vec{v}_{s}  | \text{e}^{-\text{i}\mathcal{L}t}|  \hat{\vec{q}}\cdot \vec{v}_{s}   \rangle.
\end{equation}
The reduction to the original dynamics  
 stems from the property $\lim_{q\to 0}\mathcal{L}\mathcal{P}_{\rho^{(s)}}=\sum_{\sigma \kappa}Q_{\sigma}^{(s)}|j^{\perp,(s)}_{\sigma}(\vec{q}\to 0)\rangle[(\mathbf{S}^{(s)})^{-1}]_{\sigma \kappa}\langle \rho^{(s)}_{\kappa}(q\to 0)|$ supplemented with the rotational symmetry around the $z$ axis $\langle\hat{\vec{q}}\cdot \vec{v}_{s} |\mathcal{L}^k| j^{\perp,(s)}_{\sigma}(q\to 0)\rangle\equiv 0$ for $k\in\mathbb{N}$.

The hierarchy of equations of motion is continued by utilizing in Eq.~\eqref{eq:master} the projector $\mathcal{P}_{j^{(s)}}=\sum_{\beta}\sum_{\kappa \lambda}  | j_{\kappa}^{\beta,(s)}(\vec{q})\rangle[(\bm{\mathcal{J}}^{(s)})^{-1}]^{\beta,(s)}_{\kappa \lambda} \langle j_{\lambda}^{\beta,(s)}(\vec{q})| $, and using the tagged-particle current density $|j_{\nu}^{\beta,(s)}(\vec{q})\rangle$ as a variable by  a simultaneous replacement of the Liouville operator $\mathcal{L}\to\mathcal{Q}_{\rho^{(s)}}\mathcal{L}\mathcal{Q}_{\rho^{(s)}}$. This yields a second Zwanzig-Mori equation of motion for the reduced tagged-particle current correlator,
\begin{align}\label{eq:eom2}
&\dot{\mathcal{K}}^{\alpha\beta,(s)}_{\mu\nu}(q,t) \nonumber \\
&+ \sum_{\kappa\lambda} \sum_{\gamma= \parallel, \perp} \int_0^t  \mathcal{J}^{\alpha,(s)}_{\mu \kappa} \mathcal{M}^{\alpha\gamma,(s)}_{\kappa\lambda}(q,t-t') \mathcal{K}^{\gamma\beta,(s)}_{\lambda\nu}(q,t') \diff t' = 0.
\end{align}
All the complexity of the complicated interactions is thereby hidden in the effective force-kernel  $\mathcal{M}^{\alpha\beta,(s)}_{\mu\nu}(q,t)=\left[\left(\bm{\mathcal{J}}^{(s)}\right){}^{-1}\bm{\mathfrak{M}}^{(s)}(q,t)\left(\bm{\mathcal{J}}^{(s)}\right){}^{-1}\right]^{\alpha\beta}_{\mu\nu}$, and it is specified by the formal expression,
\begin{equation}\label{eq:memory}
\mathfrak{M}^{\alpha \beta,(s)}_{\mu\nu}(q,t)=\langle \mathcal{L} j_{\mu}^{\alpha, (s)}(\vec{q})|\mathcal{Q}^{(s)}e^{-i \mathcal{L}'^{(s)}t}\mathcal{Q}^{(s)}|\mathcal{L}j_{\nu}^{\beta, (s)}(\vec{q}) \rangle.
\end{equation}
The time evolution of the kernel is provided by the reduced Liouville operator $\mathcal{L}'^{(s)}=\mathcal{Q}^{(s)}\mathcal{L}\mathcal{Q}^{(s)}$ decorated by the orthogonal projector  $\mathcal{Q}^{(s)}=\bm{1}-\mathcal{P}_{\rho^{(s)}}-\mathcal{P}_{j^{(s)}}$. It  satisfies again  the symmetry properties found for the tagged-particle current correlators; in particular it decouples in the limit $q\to0$: $\lim_{q\to 0}\mathcal{M}^{\alpha \beta,(s)}_{\mu\nu}(q,t)=:\delta^{\alpha \beta}\mathcal{M}^{\alpha,(s)}_{\mu\nu}(q,t)$.

The Zwanzig-Mori equations of motion can be formally solved by performing a Laplace transform, where the following convention is adopted:
\begin{equation}
\hat{S}_{\mu\nu}^{(s)}(q,z)=\text{i} \int_{0}^{\infty}\diff t \, S_{\mu\nu}^{(s)}(q,t) \exp (\text{i} zt ), \qquad \Imag[z]>0.
\end{equation}
Then, in the Laplace domain the time-correlation functions are represented by matrix elements of the resolvent operator $\hat{\mathcal{R}}(z) =({\cal L} - z)^{-1}$. The first Zwanzig-Mori equation, Eq.~\eqref{eq:eom1}, yields a matrix equation for the incoherent scattering function:
\begin{equation}\label{eq:MZ2}
\hat{\mathbf{S}}^{(s)}(q,z) = -\left[ z \left(\mathbf{S}^{(s)}\right){}^{-1} +\left(\mathbf{S}^{(s)}\right){}^{-1}\hat{\mathbf{K}}^{(s)}(q,z) \left(\mathbf{S}^{(s)}\right){}^{-1}\right]{}^{-1},
\end{equation}
where a matrix notation $[\hat{\bm{S}}^{(s)}(q,z)]_{\mu\nu}=\hat{S}^{(s)}_{\mu\nu}(q,z)$ has been introduced. By linearity, Eq.~\eqref{eq:contract_t} retains its shape in the Laplace domain,
\begin{equation}\label{eq:contract}
 \hat{K}_{\mu\nu}^{(s)}(q,z)= \sum_{\alpha\beta=\parallel,\perp}b^{\alpha}(q,Q_{\mu}^{(s)}) \hat{\mathcal{K}}^{\alpha\beta,(s)}_{\mu\nu}(q,z)
 b^{\beta}(q,Q_{\nu}^{(s)}),
\end{equation}
and finally, the reduced tagged-particle current correlator is expressed by inversion in terms of the tagged-particle correlator of the forces,
\begin{equation}\label{eq:MZ1}
\hat{\bm{\mathcal K}}{}^{(s)}(q,z)= -\left[ z (\bm{\mathcal J}^{(s)})^{-1} +  \hat{\bm{\mathcal M}}{}^{(s)}(q,z)\right]{}^{-1}.
\end{equation}

The short-time dynamics ($t\to 0$) of the tagged-particle correlators follows upon expansion of the Zwanzig-Mori equations in the Laplace domain for high frequencies. Assume the asymptotic behavior of the force correlator to be $\bm{\mathcal{M}}^{(s)}(q,z)=\mathcal{O}(z^{-1})$ for $z\to \infty $, one readily infers
\begin{align}
&\hat{\mathbf{ K}}^{(s)}(q,z)= -z^{-1} \mathbf{J}^{(s)}(q) + \mathcal{O}(z^{-3}),\nonumber \\
&\hat{\mathbf{S}}^{(s)}(q,z) = -z^{-1} \mathbf{S}^{(s)} - z^{-3}\mathbf{J}^{(s)}+\mathcal{O}(z^{-5}),
\end{align}
where $[\mathbf{J}^{(s)}]_{\mu\nu}(q)=\sum_{\alpha}b^{\alpha}(q,Q^{(s)}_{\mu})\mathcal{J}^{\alpha,(s)}_{\mu\nu}b^{\alpha}(q,Q^{(s)}_{\nu})$. In the time domain one obtains for the incoherent scattering function the behavior for short times ($t \to 0$),
\begin{equation}
\mathbf{S}^{(s)}(q,t)=\mathbf{S}^{(s)}-\mathbf{J}^{(s)}(q) t^2/2 +\mathcal{O}(t^4).
\end{equation}
Memory effects enter the equation
 at the earliest within order $\mathcal{O}(t^4)$. For the in-plane dynamics $S^{(s)}_{00}(q,t)=1- (q v^{(s)}_{\text{th}})^2 t^2/2 +\mathcal{O}(t^4)$ one recovers the conventional ballistic motion along the plates for times $\mathcal{O}(t^2)$.
The Zwanzig-Mori equations of motion are formally exact and can serve as a starting point for suitable approximations. The mode-coupling theory deals with a self-consistent ansatz to describe the slow motion of the tagged particle mediated by the cage effect.
\subsection{Mode-coupling theory}
Transport of dense liquids is slowed down drastically  due to the strong mutual interactions of the particles. A self-consistent description of the local caging forces is provided by the MCT~\cite{Goetze:Complex_Dynamics}, where the \emph{a priori} unknown force correlator is represented by a functional of the collective intermediate scattering function. For the description of the dynamics of a tagged particle the functional is coupled to the collective intermediate scattering function of the host liquid and linearly to the incoherent scattering function itself~\cite{Goetze:Complex_Dynamics,Fuchs:1998}. Here we provide the MCT  ansatz (elaborated in detail in Appendix~\ref{sec:Overlap}) for the kernel $\bm{\mathcal{M}}^{(s)}(q,t)$ containing the complicated forces to describe the tagged-particle dynamics in confined geometry, which is explicitly formalized to
\begin{align}\label{eq:effective}
&\mathcal{M}^{\alpha\beta,(s)}_{\mu\nu}(q,t)\approx\mathcal{F}_{\mu\nu}^{\alpha\beta,(s)}[\mathbf{S}(t),\mathbf{S}^{(s)}(t);q] \nonumber \\
=&\frac{1}{N} \sum_{\vec{q}_{1},\vec{q}_{2}=\vec{q}-\vec{q}_{1}}\sum_{\substack{\mu_{1}\mu_{2}\\ \nu_{1} \nu_{2}}}\mathcal{Y}^{\alpha,(s)}_{\mu,\mu_{1}\mu_{2}}(\vec{q},\vec{q}_{1}\vec{q}_{2})\nonumber\\
&\times S_{\mu_{1}\nu_{1}}(q_{1},t) S^{(s)}_{\mu_{2}\nu_{2}}(q_{2},t) \mathcal{Y}^{\beta,(s)}_{\nu,\nu_{1}\nu_{2}}(\vec{q},\vec{q}_{1}\vec{q}_{2})^*,
\end{align}
with corresponding vertices,
\begin{align}\label{eq:Ypsilon}
\mathcal{Y}^{\alpha,(s)}_{\mu,\mu_{1}\mu_{2}}(\vec{q},\vec{q}_{1}\vec{q}_{2})=&-\frac{n_{0}}{ L L_{s}} \delta_{\vec{q},\vec{q}_{1}+\vec{q}_{2}}\sum_{\sigma} [(\mathbf{S}^{(s)})^{-1}]_{\mu \sigma}\nonumber\\
&\times b^{\alpha}(\hat{\vec{q}}\cdot \vec{q}_{1},Q_{\sigma-\mu_{2}}^{(s)}) c_{\sigma-\mu_{2},\mu_{1}}^{(s)}(q_{1}).
\end{align}
The vertices are characterized by the generalized direct correlation function $c^{(s)}_{\mu\nu}(q)$ which quantifies the static interaction of the tagged particle with the complex surrounding confined liquid. The direct correlation function has been introduced by adapting the Ornstein-Zernike equation to the confining geometry; see Appendix~\ref{sec:Ornstein} and explicitly Eq.~\eqref{eq:OZ_tagged}. It represents a ``mixed'' correlator where the first index $\sigma$ refers to the commensurable mode of the tagged particle $Q^{(s)}_{\sigma}=2\pi \sigma/L_{s}$ and $\tau$ to the commensurable mode for the host-liquid particles $Q_{\tau}=2\pi\tau/L$ (see Appendix~\ref{sec:Ornstein}). The MCT functional for the tagged-particle dynamics neatly formalizes the coupling to the  surrounding liquid by its dependence on the coherent scattering function $S_{\mu\nu}(q,t)=N^{-1}\langle \rho_{\mu}(\vec{q},t)| \rho_{\nu}(\vec{q})\rangle$.

One finds that in the long-wavelength limit $q \to 0$ the mode-coupling functional of the tagged particle, Eq.~\eqref{eq:effective}, decouples due to $\sum_{\vec{q}_{1}}\hat{\vec{q}}\cdot \vec{q}_{1}\equiv 0$ with respect to the channel indices
\begin{equation}
\mathcal{M}_{\mu \nu}^{\alpha \beta,(s)}(q\to 0,t)=:\delta^{\alpha \beta} \mathcal{M}_{\mu \nu}^{\alpha,(s)}(t),
\end{equation}
where the thermodynamic limit $N,A \to \infty$ at constant $n_0=N/A$ has been anticipated. Explicitly, the tagged-particle correlator becomes
\begin{align}\label{eq:tagged_hyd}
\mathcal{M}_{\mu \nu}^{\alpha,(s)}(t)&= n_{0}\int_{0}^{\infty} \diff k k \sum_{\substack{\mu_{1}\mu_{2}\\ \nu_{1} \nu_{2}}} y^{\alpha,(s)}_{\mu,\mu_{1}\mu_{2}}(k) \nonumber \\
&\times S_{\mu_{1}\nu_{1}}(k,t) S^{(s)}_{\mu_{2}\nu_{2}}(k,t) y^{\alpha,(s)}_{\nu,\nu_{1}\nu_{2}}(k)^*.
\end{align}
with corresponding vertices
\begin{align}
y^{\alpha,(s)}_{\mu,\mu_{1}\mu_{2}}(k)=&\frac{1}{\sqrt{2 \pi}  L L_{s}}\sum_{\sigma}[(\mathbf{S}^{(s)})^{-1}]_{\mu \sigma}\nonumber\\
&\times b^{\alpha}(k/\sqrt{2},Q^{(s)}_{\sigma-\mu_{2}}) c_{\sigma-\mu_{2},\mu_{1}}^{(s)}(k).
\end{align}
The decoupling property $\mathcal{M}_{\mu \nu}^{\alpha \beta,(s)}(q\to 0,t)=:\delta^{\alpha \beta} \mathcal{M}_{\mu \nu}^{\alpha,(s)}(t)$ is therefore respected by the theory, as required by rotational symmetry around the $z$ axis [see discussion of Eq.~\eqref{eq:memory}]. As a consequence, the reduced tagged-particle current correlator of the MCT inherits the decoupling property as well, $\mathcal{K}^{\alpha \beta,(s)}_{\mu \nu}(q\to 0,t)=\delta^{\alpha \beta}\mathcal{K}_{\mu\nu}^{\alpha,(s)}(q\to 0,t)$ [see Eq.~\eqref{eq:eom2}].

\subsection{Incoherent glass-form factors}

Generically, the MCT ansatz entails a bifurcation scenario for the long-time limit of the incoherent scattering function~\cite{Goetze:Complex_Dynamics,Fuchs:1998}, which is also referred to as the incoherent nonergodicity parameter or glass-form factors. This parameter,
\begin{equation}
F^{(s)}_{\mu\nu}(q):=\lim_{t\rightarrow\infty}S^{(s)}_{\mu\nu}(q,t),
\end{equation}
enables one to distinguish between an ergodic liquid state ($F^{(s)}_{\mu\nu}(q)\equiv 0$) and an arrested localized state $F^{(s)}_{\mu\nu}(q)\neq 0$. For a given long-time limit of the collective dynamics
$F_{\mu\nu}(q):=\lim_{t\rightarrow\infty}S_{\mu\nu}(q,t)$, the frozen-in part of the fluctuating forces is obtained as
\begin{equation}\label{eq:N}
\bm{\mathcal{F}}^{(s)}[\mathbf{F},\mathbf{F}^{(s)};q]:= \bm{\mathcal{M}}^{(s)}(q,t\to \infty ).
\end{equation}
With
\begin{align}
&[(\mathbf{N}^{(s)})^{-1}]_{\mu\nu}(q)\nonumber\\
=& \sum_{\alpha\beta =\parallel,\perp} b^\alpha(q,Q^{(s)}_\mu)
[\bm{\mathcal{F}}^{(s)}[\mathbf{F},\mathbf{F}^{(s)};q]^{-1}]_{\mu\nu}^{\alpha\beta}
b^\beta(q,Q^{(s)}_\nu),
\end{align}
one infers from the  Zwanzig-Mori equation of motion, Eq.~\eqref{eq:MZ2}, a self-consistent equation for the incoherent nonergodicity parameter,
\begin{align}\label{eq:F}
\mathbf{F}^{(s)}(q) &= \left[ \left(\mathbf{S}^{(s)}\right){}^{-1} + \left(\mathbf{S}^{(s)}\right){}^{-1} (\mathbf{N}^{(s)})^{-1}(q) \left(\mathbf{S}^{(s)}\right){}^{-1} \right]{}^{-1}\nonumber\\
&=\mathbf{S}^{(s)}- \left[ \left(\mathbf{S}^{(s)}\right)^{-1}+ \mathbf{N}^{(s)}(q)\right]{}^{-1}.
\end{align}
To avoid cumbersome notation we allow $[\mathbf{N}^{(s)}]^{-1}(q)$ to become formally infinite, and in this case we take $\mathbf{N}^{(s)}(q)=0$. The equation for the nonergodicity parameter has many solutions, in particular $\mathbf{F}^{(s)}(q) \equiv 0$ represents the  trivial solution which corresponds to the liquid state.

\subsection{Properties of the solutions}

The closed set of equations for the incoherent scattering function and for the self-nonergodicity parameter in confined geometry belongs to a certain class of MCT-equations, whose properties have been recently investigated~\cite{Lang:2013}. This class is characterized by multiple relaxation channels as found, for instance, for molecular liquids, where the local density varies due to rotation and translation of the molecule~\cite{Franosch:1997,Scheidsteger:1997}. In the present context of the motion of a tagged particle in a dense liquid confined between parallel walls the two currents arise due to the parallel and perpendicular momentum with respect to the surfaces [see Eq.~\eqref{eq:continuity}]. Since the coherent dynamics of the embedding liquid~\cite{Lang:2010,Lang:2012} is assumed to be known [Eq.~\eqref{eq:effective}], the MCT functional for the tagged particle is a \textit{linear} functional in the incoherent scattering function $\bm{S}^{(s)}(q,t)$. It is literally straightforward to make contact with Ref.~\cite{Lang:2013}; one merely has to show, that the tagged-particle MCT functional preserves the same properties as the MCT-functional for the collective dynamics~\cite{Lang:2012}. To avoid a cumbersome repetition of the same arguments as presented in Ref.~\cite{Lang:2013}, we merely report the conclusions relevant for the tagged-particle motion.

The tagged-particle MCT functional maps non-negative matrices $\bm{F}^{(s)}(q) \succeq 0$ in the mode indices $\mu,\nu$ to non-negative matrices with respect to the double indices $\gamma := (\alpha,\mu), \delta:= (\beta,\nu)$ for each wave number $q$. Generically, all vertices are nonvanishing and all components of the functional are positive matrices $\bm{\mathcal{F}}^{(s)}[\mathbf{F} ,\mathbf{F}^{(s)};q ] \succ 0$ provided the arguments are positive $\mathbf{F}^{(s)}(q) \succ 0$. By linearity, the tagged-particle MCT functional preserves order $\bm{\mathcal{F}}^{(s)}[\mathbf{F},\mathbf{F}^{(s)};q]-\bm{\mathcal{F}}^{(s)}[\mathbf{F},\mathbf{E}^{(s)};q] \succeq 0$, if $\mathbf{F}^{(s)}(q)-\mathbf{E}^{(s)}(q)\succeq 0$ for each $q$.

Matrix valued correlation functions $S^{(s)}_{\mu \nu}(q,t)$ satisfy the condition  $\sum_{ij}\sum_{\mu \nu}(\xi_{i}a_{\mu})^* S^{(s)}_{\mu \nu}(q,t_{i}-t_{j})(\xi_{j}a_{\nu})\geq 0$~\cite{Goetze:Complex_Dynamics} for any finite set of times $t_i\in \mathbb{R}$ and complex numbers $\xi_{i},a_{\mu} \in \mathbb{C}$. The MCT functional of the tagged particle indeed preserves this property and therefore correlation functions are mapped onto correlation functions. In this sense it satisfies the same properties as the MCT functional of the collective dynamics. Since the structure of the Zwanzig-Mori equations is of identical form, the general properties of the solutions $\bm{S}^{(s)}(q,t)$ share the same properties as the solutions for the collective dynamics $\bm{S}(q,t)$ and all conclusions from Ref.~\cite{Lang:2013} can be taken over. The solutions are unique and can be obtained by a convergent iteration scheme. Similarly, the solution of the self-consistent equation for the self-nonergodicity parameter [see Eq.~\eqref{eq:N}] is the maximum of all possible solutions, which is referred to as the maximum theorem. Provided the limit $\lim_{t \to \infty}\bm{S}^{(s)}(q,t)$ exists, it coincides with the maximum solution of the self-consistent equation for the self-nonergodicity parameter, which is a consequence of a generalized covariance principle demonstrated in Ref.~\cite{Lang:2013}.

To mimic overdamped Brownian dynamics one performs in Eq.~\eqref{eq:MZ1} the substitution,
\begin{align}\label{eq:OD}
&\left[ z (\bm{\mathcal J}^{(s)}){}^{-1} +  \hat{\bm{\mathcal M}}{}^{(s)}(q,z)\right]{}^{-1}\nonumber\\
&\to \left[ \text{i} \left(\bm{\mathcal{D}}^{(s)}\right){}^{-1}(q) + \hat{\bm{\mathcal M}}{}^{(s)}(q,z)\right]{}^{-1}
\end{align}
with a positive definite matrix $\bm{\mathcal{D}}^{(s)}(q)\succ 0$ characterizing the short-time diffusion of the tracer. In this case, following similar arguments as elaborated in Ref.~\cite{Lang:2013}, the solutions of the MCT-equation for $\mathbf{S}^{(s)}(q,t)$ belong to a subclass of correlation functions, which are purely relaxational~\cite{Lang:2013}.

\section{In-plane dynamics of the tagged particle}
Monitoring the confined liquid from a top-view perspective one probes solely the lateral coordinates of the particles confined within the slab. Then, the system appears as a two-dimensional projection, but its dynamics strictly depends on the wall separation. The in plane-dynamics is encoded in the matrix-element $S_{00}^{(s)}(q,t)=\langle \text{e}^{-\text{i} \vec{q} \cdot \left[\vec{r}_s(t)-\vec{r}_s(0)\right]} \rangle$ of the generalized incoherent scattering function. To extract the in-plane dynamics from the matrix-valued theory we implement a novel projection-operator technique with nonorthogonal projectors; see Appendix~\ref{sec:Resolvent}.
Thereby it is demonstrated that in the Laplace domain the in-plane density-fluctuations can be represented as a generalized diffusion equation,

\begin{equation}\label{eq:diffusion}
 \hat{S}^{(s)}_{00}(q,z)=\frac{-1}{z+q^2 \hat{D}(q,z)},
\end{equation}
characterized by the frequency-dependent diffusion coefficient $\hat{D}(q,z)$ (see Appendix~\ref{sec:Resolvent}). Explicitly, it is given by the current correlators of the tagged particle
\begin{align}\label{eq:Diffusion}
\hat{D}(q,z)=&\hat{K}^{(s)}_{00}(q,z)/q^2\nonumber\\
&+ \sum_{\nu}\hat{K}^{(s)}_{0\nu}(q,z)\Big(\left[ z\bm{1} +\mathcal{Q}\hat{\bm{K}}^{(s)}(q,z)\right]{}^{-1}\nonumber\\
&\times\mathcal{Q}\hat{\bm{K}}^{(s)}(q,z)\Big){}_{\nu 0}/q^2,
\end{align}
with $\mathcal{Q}=[\bm{S}^{(s)}]^{-1}-\mathcal{P}$ and $[\mathcal{P}]_{\mu\nu}=\delta_{\mu 0} [1/S_{00}^{(s)}] \delta_{\nu 0} = \delta_{\mu 0}  \delta_{\nu 0}$; see for a derivation Appendix~\ref{sec:Resolvent}. While the first term in Eq.~\eqref{eq:Diffusion} is of order $\mathcal{O}(q^{0})$, the second term vanishes in the long-wave-length limit as can be inferred from  Eq.~\eqref{eq:contract} and by utilizing the decoupling property $\lim_{q\to 0}\hat{\mathcal{K}}^{\parallel \perp,(s)}_{0 \mu}(q\to 0,z)\equiv 0$.
Expanding Eq.~\eqref{eq:diffusion} in powers of the wave number $q$ one obtains
\begin{equation}
 \hat{S}^{(s)}_{00}(q,z)=-\frac{1}{z}+ \frac{q^2 \hat{Z}_{\parallel}(z)}{z^2}+\mathcal{O}(q^4),
\end{equation}
where $\hat{Z}_{\parallel}(z)=\lim_{q \to 0}\hat{D}(q,z)$ represents the long-wave-length  limit of the frequency-dependent diffusion coefficient. From Eq.~\eqref{eq:Diffusion} one infers the simple relation,
\begin{align}
&\hat{Z}_{\parallel}(z)=\hat{\mathcal{K}}^{\parallel \parallel,(s)}_{00}(q\to 0,z).
\end{align}

A similar expansion of the incoherent scattering function $S_{00}^{(s)}(q,t)=\langle \text{e}^{-\text{i} \vec{q} \cdot \left[\vec{r}_s(t)-\vec{r}_s(0)\right]} \rangle$ with respect to the wave number $q$ yields
\begin{equation}
S_{00}^{(s)}(q,t)= 1 - q^2 \delta r^{2}_{\parallel}(t)/4+\mathcal{O}(q^4),
\end{equation}
where the in-planar mean-square displacement $\delta r^{2}_{\parallel}(t)=\langle \left[\vec{r}_s(t)-\vec{r}_s(0) \right]^2 \rangle$ occurs. Then,  one identifies $\hat{Z}_{\parallel}(z)$ to be merely the Laplace transform of the in-planar mean-square displacement
\begin{align}\label{eq:spect}
\hat{Z}_{\parallel}(z)&=-\text{i} \left(z^2/4 \right) \int_{0}^{\infty}\text{e}^{\text{i} z t} \delta r^{2}_{\parallel}(t)  \diff t \\ \nonumber
&=\text{i} \int_{0}^{\infty} \text{e}^{\text{i} z t} Z_{\parallel}(t) \diff t .
\end{align}
Integrating  by parts twice yields the relation
\begin{align}
 Z_{\parallel}(t)=\frac{1}{4}\delta \ddot{r}_{\parallel}^2(t)= \frac{1}{2} \langle \vec{v}_s(t)\cdot \vec{v}_s(0)  \rangle,
\end{align}
and one concludes that  $Z_{\parallel}(t)$ is the in-planar velocity-velocity autocorrelation function of the tagged particle and can be inferred from Eq.~\eqref{eq:VACF}. By the Green-Kubo relation the in-planar long-time diffusion coefficient is given as the integrated velocity-velocity auto-correlation function
\begin{equation}
D_{\parallel}= \int_{0}^{\infty}  Z_{\parallel}(t) \diff t,
\end{equation}
and from Eq.~\eqref{eq:spect} it can be calculated from the Laplace transformed velocity-autocorrelation function in the limit of small frequencies,
\begin{align}\label{eq:microD}
D_{\parallel}&= \lim_{z \to 0} \hat{Z}_{\parallel}(z)/\text{i}\nonumber \\
&=\left(\left[\int_{0}^{\infty} \bm{\mathcal{M}}^{\parallel,(s)}(t)\diff t \right]^{-1}\right){}_{00}
\end{align}
where $[\bm{\mathcal{M}}^{\parallel,(s)}(t)]_{\mu\nu}=\mathcal{M}^{\parallel,(s)}_{\mu\nu}(t)$ from Eq.~\eqref{eq:tagged_hyd}. Evaluating the kernel for all times, the diffusion coefficient parallel to the walls is given by the derived formula. For a particular interaction of the tagged particle with the host-liquid particles, it is a function of the wall separation and the average density of the solvent $D_{\parallel}\equiv D_{\parallel}(L,L_{s},n_{0})$. The feature of a localization transition can be read: The diffusion coefficient is zero, $D_{\parallel}\equiv 0$ in the case of a localized dynamics $ \lim_{t \to \infty}\bm{\mathcal{M}}^{\parallel,(s)}(t)\succ 0$ see Eq.~\eqref{eq:tagged_hyd}), or else the diffusion coefficient is finite $0<D_{\parallel}<\infty$ and quantifies the rapidity of lateral particle transport in the confined environment. The formula, Eq.~\eqref{eq:microD}, represents a microscopic expression for the diffusion coefficient and we anticipate a nontrivial dependence as a function of the wall separation, as it has been elucidated by recent computer simulations~\cite{Mittal:2008}.

Similarly, a measure of the cage size characterizing the localized state is the localization length, which is anisotropic within the confined geometry. The isotropic part is determined by the in-plane projection,
\begin{equation}
\lim_{t\to \infty}\delta r^{2}_{\parallel}(t)=4 \ell_{\parallel}^2.
\end{equation}
By  Laplace transform of the in-planar mean-square displacement together  with Eq.~\eqref{eq:spect} and Eq.~\eqref{eq:MZ1} the localization length is obtained by the long-time limit of the tagged-particle MCT functional,
\begin{equation}
\ell_{\parallel}=\left[\left(\left[\lim_{t \to \infty}\bm{\mathcal{M}}^{\parallel,(s)}(t)\right]^{-1}\right){}_{00}\right]^{1/2},
\end{equation}
which is finite beyond a certain critical point when the forces freeze in.
\section{Summary and Conclusions}
We have developed a theory for the dynamics of an arbitrarily sized tagged particle moving in a densely packed confined liquid. Formally exact Zwanzig-Mori equations for the incoherent scattering function have been derived by the projection operator formalism and closed by a self-consistent mode-coupling ansatz. The incoherent scattering function is decomposed into commensurate plane waves which exactly match the gap between the adjacent flat walls. The MCT functional is a \emph{linear} functional in the incoherent scattering function and vertices control the coupling strength to the dynamics of the surrounding fluid. The nontrivial spatial information of the interactions with the solvent is encoded in a generalized direct-correlation function, only. Thereby, we have adapted the  Ornstein-Zernike equation to inhomogeneous liquids to describe the static coupling of a tagged particle with the surrounding host liquid.

The MCT equations of motion for the tagged particle in confined geometry belong to a particular class of mode-coupling equations, which are characterized by multiple relaxation channels and identified in the present work as the current of the tagged particle parallel and perpendicular to the walls. Since the tagged-particle MCT functional displays similar properties as the coherent MCT functional, all conclusions of Ref.~\cite{Lang:2013} can be taken over, e.g., the solutions satisfy the constraints set by probability theory, are unique, and can be obtained by a convergent iteration. Furthermore, a covariance principle holds, which guarantees that the  maximum solution of the self-consistent equation for the self-nonergodiciy parameter coincides with the long-time limit of the incoherent scattering function ~\cite{Lang:2013}.

The generalized incoherent scattering function contains all the spatial information on the dynamics of the tagged particle within the commensurable modes. Thereby we have defined and provided the relevant physical quantities in confined geometry, which should be addressed in experiments and computer simulations for the tagged-particle motion between parallel walls.

Monitoring the confined liquid from a top-view perspective one probes solely the lateral coordinates of the particles. The dynamics of these coordinates has been addressed in experiments~\cite{Nugent:2007} as well as in computer simulations~\cite{Mittal:2008}. In Ref.~\cite{Mittal:2008} the corresponding diffusion coefficient parallel to the walls has been extracted and oscillatory behavior as a function of the wall separation detected~\cite{Mittal:2008}. In the present work the in-plane dynamics is encoded in the generalized incoherent scattering function as the matrix element $S_{00}^{(s)}(q,t)$. Using the resolvent calculus, we have derived an equation of motion for this particular matrix element. We have performed the long-wave-length limit to enter the generalized hydrodynamic regime  and via a Green-Kubo relation a microscopic expression for the in-planar diffusion coefficient has been derived.

An interesting regime is when the distance between the plates becomes  small~\cite{Franosch:2012,*Lang:2014}. Assume the tagged particle to be smaller in extension  $L_{s}\geq L$ than the embedding solvent, the limit $L\to 0$ provides a system, where the motion of the solvent is effectively two-dimensional~\cite{Bayer:2007,Hajnal:2009} while the tagged-particle motion still is enabled to perform fluctuations in the $z$ direction. Vice versa for $L_{s}\leq L$ the limit $L_{s}\to 0$ is anticipated to describe a two-dimensional (2D) motion of the tracer in a solvent with an inhomogeneous structure perpendicular to the walls. An analytical access to these regimes requires primarily the behavior of the structural quantities in the 2D limit.

We have set the analytical fundament for the study of the dynamics of a tagged particle in confined geometry where the host liquid is densely packed. Future numerical implementations of the derived equations of motion are highly desirable to gain a deeper understanding of the dynamical processes of particle transport in such environments  where packing effects play a major role. Interesting dynamical phenomena are expected, which are attributed to incommensurability effects in the confined geometry and an interesting interplay of the tagged particle with its confined liquid is anticipated.

Generically, one expects that a structural arrest of the host liquid implies a localization of the tagged particle. However, for weak coupling to the embedding liquid, e.g. diminishing the particle size or interaction strength with the host, the tracer may become mobile again and is then enabled to meander through the frozen disordered array. Such a critical point marks a so-called type-A glass transition characterized by a continuous change within the incoherent glass-form factor at a critical point~\cite{Franosch:1994}.

Quite generally the glass transition as well as the diffusion-localization phenomena become more involved for  confined geometry, where the incoherent glass-form factor crucially depends on the wall separation. The tracer may be utilized as a sensor to probe the structure of the glass by determining the localization length. Furthermore, confinement-induced localization effects are conceivable, where by pure variation of the wall separation and fixed density of the host liquid the tagged-particle dynamics becomes localized and diffusive again. One may as well speculate the existence of periodic traversing of percolation edges  upon variation of the  wall separation in the case of weak coupling to the solvent.


\appendix

\section{Tagged-particle MCT-functional in confined geometry}
\label{sec:Overlap}
In this appendix the MCT-functional for  tagged particle motion is  constructed in detail from the formal expression,
\begin{equation}\label{eq:memoryapp}
\mathfrak{M}^{\alpha \beta,(s)}_{\mu\nu}(q,t)=\langle \mathcal{L} j_{\mu}^{\alpha, (s)}(\vec{q})|\mathcal{Q}^{(s)}e^{-i \mathcal{L}'^{(s)}t}\mathcal{Q}^{(s)}|\mathcal{L}j_{\nu}^{\beta, (s)}(\vec{q}) \rangle.
\end{equation}
The strategy is to project the forces onto the product of the tagged particle and collective density $\delta\rho_{\mu}(\vec{q}_{1})\delta\rho^{(s)}_{\nu}(\vec{q}_{2})$~\cite{Goetze:Complex_Dynamics}, where the collective density is denoted by $\rho_{\mu}(\vec{q})=\sum_{n=1}^{N}\text{e}^{\text{i}\vec{q}\cdot\vec{r}_{n}}\exp[\text{i}Q_{\mu} z_{n}]$ with commensurable modes for the host-liquid particles $Q_{\mu}=2 \pi \mu/L,\mu \in\mathbb{Z}$~\cite{Lang:2012}. Then, a successive factorization of the four-point correlation function  into a product of density correlation functions is employed. The projector in compact notation $i=(\vec{q}_{i},\mu_{i})$ and $i^{\prime}=(\vec{q}_{i}^{\prime},\mu_{i}^{\prime})$ (see also Ref.~\cite{Scheidsteger:1997}) is given by
\begin{equation}
 \mathcal{P}_{\rho\rho^{(s)}}=\sum_{11'22'} |\delta \rho(1)\delta \rho(2)^{(s)}\rangle \mathfrak{g}^{(s)}(12;1'2') \langle \delta \rho(1') \delta\rho^{(s)}(2')|,
 \end{equation}
and is inserted twice to sandwich the reduced backwards-time evolution operator ${\cal R}'{}^{(s)}$  in the force kernel $\mathfrak{M}^{\alpha \beta ,(s)}_{\mu \nu}(q,t)$ of Eq.~\eqref{eq:memoryapp}. The matrix $\mathfrak{g}^{(s)}(12;1'2')$  ensures idempotency of the projector, $\mathcal{P}_{\rho\rho^{(s)}}^2 = \mathcal{P}_{\rho\rho^{(s)}}$.
 A subsequent factorization and reduction to the original dynamics yields a product of the form
\begin{align}\label{eq:factor}
\langle\delta\rho(1)^*\delta\rho^{(s)}&(2)^*\exp\left(-i \mathcal{L}'^{(s)}t\right)\delta \rho(1^{\prime})\delta\rho^{(s)}(2^{\prime})\rangle\nonumber\\
&\approx N S(1,1^{\prime},t) S^{(s)}(2,2^{\prime},t),
\end{align}
where the generalized coherent intermediate scattering function of the host liquid $[\bm{S}(q,t)]_{\mu\nu}=N^{-1}\langle \rho_{\mu}(\vec{q},t)|\rho_{\nu}(\vec{q})\rangle$~\cite{Lang:2012} has been introduced. The combination $(1' \leftrightarrow 2')$ is of order $\mathcal{O}(N^0)$ and therefore of negligible weight in the thermodynamic limit. From Eq.~\eqref{eq:factor} one infers to initial time $t=0$ the corresponding static factorization of the four-point correlator,
\begin{align}
\langle \delta\rho(1)^*&\delta\rho^{(s)}(2)^* \delta\rho(1^{\prime})\delta\rho^{(s)}(2^{\prime})\rangle\nonumber\\
\approx &N S(1,1^{\prime}) S^{(s)}(2,2^{\prime}).
\end{align}
The preceding  manipulations suggest employing the consistent factorization of the normalization matrix,
\begin{equation}
 \mathfrak{g}(12;1^{\prime}2^{\prime}) \approx N^{-1} \mathbf {S}^{-1}(1,1^{\prime})\left(\mathbf {S}^{(s)}\right){}^{-1}(2,2^{\prime}).
\end{equation}
As a consequence, the tagged-particle memory kernel is expressed  in terms of the incoherent and coherent intermediate scattering function of the host liquid,
\begin{align}\label{eq:Memory}
\mathfrak{M}^{\alpha\beta,(s)}_{\mu\nu}(q,t)\approx & \frac{1}{N} \sum_{\vec{q}_{1},\vec{q}_{2}=\vec{q}-\vec{q}_{1}}\sum_{\substack{\mu_{1}\mu_{2}\\ \nu_{1} \nu_{2}}}\mathcal{X}^{\alpha,(s)}_{\mu,\mu_{1}\mu_{2}}(\vec{q},\vec{q}_{1}\vec{q}_{2})\nonumber\\
&\times S_{\mu_{1}\nu_{1}}(q_{1},t) S^{(s)}_{\mu_{2}\nu_{2}}(q_{2},t) \mathcal{X}^{\beta,(s)}_{\nu,\nu_{1}\nu_{2}}(\vec{q},\vec{q}_{1}\vec{q}_{2})^*.
\end{align}
Here, translational invariance parallel to the walls implies the selection rule $\vec{q}=\vec{q}_{1}+\vec{q}_{2}$ and the coupling to the time-correlation functions  is provided by the vertices,
\begin{align}
\mathcal{X}^{\alpha,(s)}_{\mu,\mu_{1}\mu_{2}}(\vec{q},\vec{q}_{1}\vec{q}_{2})=&\sum_{\kappa \lambda}\langle \mathcal{L} j_{\mu}^{\alpha,(s)}(\vec{q})^* \mathcal{Q}^{(s)}\left[\delta  \rho_{\kappa}(\vec{q}_{1})\delta  \rho_{\lambda}^{(s)}(\vec{q}_{2})\right] \rangle  \nonumber \\
&\times [\bm{S}^{-1}(q_{1})]_{\kappa \mu_{1}}   \left[\left(\bm{S}^{(s)}\right){}^{-1}\right]_{\lambda\mu_{2}},
\end{align}
and are solely determined by static equilibrium correlation functions. The overlap $\langle \mathcal{L} j_{\mu}^{\alpha,(s)}(\vec{q})^*\mathcal{Q}^{(s)}\left[\delta  \rho_{\kappa}(\vec{q}_{1})\delta  \rho_{\lambda}^{(s)}(\vec{q}_{2})\right] \rangle$ can be calculated explicitly in terms of suitable static correlation functions.  The orthogonal projector $\mathcal{Q}^{(s)}=\bm{1}-\mathcal{P}^{(s)}=\bm{1}-\mathcal{P}_{\rho^{(s)}}-\mathcal{P}_{j^{(s)}}$ is specified by the projectors $\mathcal{P}_{j^{(s)}}:=\sum_{\beta}\sum_{\sigma \tau}  | j_{\sigma}^{\beta,(s)}(\vec{q})\rangle [(\bm{\mathcal{J}}^{(s)})^{-1}]^{\beta,(s)}_{\sigma \tau} \langle j_{\tau}^{\beta,(s)}(\vec{q})|$ and $\mathcal{P}_{\rho^{(s)}}=\sum_{\sigma \tau}|\rho_{\sigma}^{(s)}(\vec{q})\rangle[(\mathbf{S}^{(s)})^{-1}]_{\sigma \tau}\langle\rho_{\tau}^{(s)}(\vec{q})|$.  Since $\mathcal{P}_{j^{(s)}}|\rho^{(s)}_{\sigma}(\vec{q})\rangle \equiv 0$ one readily infers
\begin{align}
\langle \mathcal{L} j_{\mu}^{\alpha,(s)}&(\vec{q})^*  \mathcal{Q}^{(s)}\left[\delta  \rho_{\kappa}(\vec{q}_{1})\delta  \rho_{\lambda}^{(s)}(\vec{q}_{2}) \right]\rangle\nonumber \\
&=\langle j_{\mu}^{\alpha,(s)}(\vec{q})^* \delta\rho_{\kappa}(\vec{q}_1)\left[\mathcal{L}\delta\rho_{\lambda}^{(s)}(\vec{q_{2}})\right]\rangle \nonumber\\
&-\langle\mathcal{L}j_{\mu}^{\alpha,(s)}(\vec{q})^*\mathcal{P}_{\rho^{(s)}} \left[\delta \rho_{\kappa}(\vec{q_{1}})\delta\rho_{\lambda}^{(s)}(\vec{q}_2)\right] \rangle.
\end{align}
Note, that the term $\langle j_{\mu}^{\alpha,(s)}(\vec{q})^* [\mathcal{L}\delta\rho_{\kappa}(\vec{q}_1)]\delta\rho_{\lambda}^{(s)}(\vec{q_{2}})]\rangle=0$ vanishes by symmetry $\langle\vec{v}_{s}\rangle=0$. Next we employ the continuity equation of the tracer particle $\mathcal{L}\rho_\lambda^{(s)} (\vec{q}_{2})=  \sum_{\alpha=\parallel,\perp} b^{\alpha}(q_{2},Q^{(s)}_{\lambda}) j_{\lambda}^{\alpha,(s)}(\vec{q}_{2})$ to evaluate the overlap to
\begin{align}\label{eq:over}
&\langle \mathcal{L} j_{\mu}^{\alpha,(s)} (\vec{q})^{*}  \mathcal{Q}^{(s)}\left[ \delta  \rho_{\kappa}(\vec{q}_{1})\delta  \rho_{\lambda}^{(s)}(\vec{q}_{2}) \right]\rangle
\nonumber \\
&=(v^{(s)}_{\text{th}})^2 \delta_{\vec{q},\vec{q}_{1}+\vec{q}_{2}} \Big[ b^{\alpha}(\hat{\vec{q}} \cdot \vec{q}_{2},Q_{\lambda}^{(s)})\langle \rho_{\mu-\lambda}^{(s)}(\vec{q}_{1})|\rho_{\kappa}(\vec{q}_{1})\rangle  \nonumber \\
&-\sum_{\sigma \tau} b^{\alpha} (q,Q_{\sigma}^{(s)}) S_{\mu\sigma}^{(s)} [(\mathbf{S}^{(s)})^{-1}]_{\sigma \tau} \langle \rho_{\tau-\lambda}^{(s)}(\vec{q}_{1})|\rho_{\kappa}(\vec{q}_{1})\rangle \Big].
\end{align}
In contrast to the collective dynamics ~\cite{Lang:2012}, the vertices  do not involve a triple-density correlator and therefore no convolution approximation is required. The static coupling of the microscopic tagged-particle  density to the density of the surrounding solvent is quantified by the overlap $\langle \rho_{\mu}^{(s)}(\vec{q})|\rho_{\nu}(\vec{q})\rangle$.  A compact expression in terms of a static measure for the interactions of the tagged-particle with the host liquid is provided by the direct correlation function $[\bm{c}^{(s)}(q)]_{\mu \nu}=c^{(s)}_{\mu\nu}(q)$. Via a suitably generalized Ornstein-Zernike equation (see Appendix~\ref{sec:Ornstein}), the density overlap is connected to the direct correlation function by
\begin{equation}
\langle \rho_{\mu}^{(s)}(\vec{q})|\rho_{\kappa}(\vec{q})\rangle=\frac{n_{0}}{LL_{s}} \sum_{\sigma \tau} S^{(s)}_{\mu\sigma}c^{(s)}_{\sigma\tau}(q)S_{\tau \kappa}(q),
\end{equation}
[see Eq.~\eqref{eq:OZ_tagged}], where $c^{(s)}_{\sigma\tau}(q)$ is referred to  as a ``mixed'' correlator where the first index $\sigma$ refers to the commensurable mode of the tagged particle $Q^{(s)}_{\sigma}=2\pi \sigma/L_{s}$ and correspondingly $\tau$ to the commensurable discrete mode of the host-liquid particles $Q_{\tau}=2\pi\tau/L$ (see Appendix~\ref{sec:Ornstein}). The equation for the overlap, Eq.~\eqref{eq:over}, can then be recast to
\begin{align}
\langle \mathcal{L} &j_{\mu}^{(s),\alpha} (\vec{q})^{*}  \mathcal{Q}^{(s)}\left[\delta  \rho_{\kappa}(\vec{q}_{1})\delta  \rho_{\lambda}^{(s)}(\vec{q}_{2})\right] \rangle \nonumber\\ 
=&-(v^{(s)}_{\text{th}})^2\frac{1}{LL_{s}}\delta_{\vec{q},\vec{q}_{1}+\vec{q}_{2}} n_{0} \sum_{\sigma\tau} b^{\alpha}(\hat{\vec{q}}\cdot \vec{q}_{1},Q_{\sigma}^{(s)})\nonumber\\
&\times S^{(s)}_{\mu-\lambda,\sigma}c_{\sigma\tau}^{(s)}(q_{1})S_{\tau \kappa}(q_{1}),
\end{align}
where the selection rule due to translational invariance $\vec{q}=\vec{q}_{1}+\vec{q}_{2}$ has been utilized. One finally concludes for the vertices,
\begin{align}
&\mathcal{X}_{\mu,\mu_{1}\mu_{2}}^{\alpha,(s)}(\vec{q},\vec{q}_{1},\vec{q}_{2})\\ \nonumber
&=-(v^{(s)}_{\text{th}})^2\frac{n_{0}}{ L L_{s}} \delta_{\vec{q},\vec{q}_{1}+\vec{q}_{2}} b^{\alpha}(\hat{\vec{q}}\cdot \vec{q}_{1},Q_{\mu-\mu_{2}}^{(s)})c_{\mu-\mu_{2},\mu_{1}}^{(s)}(q_{1}),
\end{align}
and the dependence on the wave number is solely encoded in terms of the direct correlation function. Decorating Eq.~\eqref{eq:Memory} by the inverse static current correlators, i.e., $\left(\bm{\mathcal{J}}^{(s)}\right){}^{-1}(\cdot) \left(\bm{\mathcal{J}}^{(s)}\right){}^{-1}$, yields then the desired expression as presented in the main text [see Eq.~\eqref{eq:effective}].

\section{Ornstein-Zernike equation}
\label{sec:Ornstein}
The spatial representation of the overlap measure $\langle \rho_{\mu}^{(s)}(\vec{q})|\rho_{\nu}(\vec{q})\rangle$ provides structural information on the arrangement of the solvent particles with respect to the tracer. We shall therefore introduce the inhomogeneous pair-distribution function $g^{(s)}(|\vec{r}-\vec{r}'|,z,z')$, which does not solely depend on the  distance between the particles, but explicitly on the transversal positions $z$ of the tracer and $z'$ of the solvent particles due to the lack of translational symmetry.  Following Ref.~\cite{Hansen:Theory_of_Simple_Liquids} we introduce the pair-distribution function by the definition
\begin{align}\label{eq:gs}
&n^{(s)}(z) g^{(s)}(|\vec{r}-\vec{r}'|,z,z')n(z')/n_{0}  \nonumber \\
&:=\sum_{n=1}^{N} \langle \delta (z-z_{s}) \delta \left[\vec{r}-\vec{r}'-(\vec{r}_{s}-\vec{r}_{n})\right] \delta(z'-z_{n}) \rangle  \nonumber \\
&=:H^{(s)}(|\vec{r}-\vec{r}'|,z,z')+n^{(s)}(z)n(z')/n_0.
\end{align}
The pair-distribution function for the solute-solvent system is related to the total correlation function by subtracting the ideal-gas part $h^{(s)}(|\vec{r}-\vec{r}'|,z,z')=g^{(s)}(|\vec{r}-\vec{r}'|,z,z')-1$, compare Ref.~\cite{Franosch:1997c,*Franosch:1999b} in the context of a molecular tracer, and one identifies $H^{(s)}(|\vec{r}-\vec{r}'|,z,z')=n^{(s)}(z) h^{(s)}(|\vec{r}-\vec{r}'|,z,z')n(z')/n_{0}$. For the Fourier transforms of ``mixed'' correlators, e.g., $X^{(s)}({|\vec{r}-\vec{r}^{\prime}|,z,z^{\prime}})$,  where the tracer positions are $(\vec{r},z)$ and solvent positions $(\vec{r}',z')$, we adopt the convention,
\begin{align} \label{eq:Fourier}
 X^{(s)}_{\mu\nu}(q)=&
 \int\limits_{-L_{s}/2}^{L_{s}/2}\!\!\!\!\diff z \!\! \int\limits_{-L/2} ^{L/2}\!\!\!\!\diff z^{\prime}\int\limits_{A} \! \diff (\vec{r}-\vec{r}') X^{(s)}({|\vec{r}-\vec{r}^{\prime}|,z,z^{\prime}})\nonumber   \\
 & \times \exp\left[- \text{i}(Q_{\mu}^{(s)}z- Q_\nu z') \right] \text{e}^{-\text{i} \vec{q} \cdot (\vec{r}-\vec{r}^{\prime})}.
\end{align}
The first mode index of the mixed correlators, here $X^{(s)}_{\mu\nu}(q)$, always refers to $Q_{\mu}^{(s)}=2\pi \mu/L_{s}$, which is the commensurable mode for the  effective slit width $L_{s}$ of the tagged particle.
One infers that the static density fluctuations of the tagged particle and the surrounding liquid can be expressed  via the relation
\begin{equation}
H^{(s)}_{\mu \nu}(q)=\langle \rho_{\mu}^{(s)}(\vec{q})|\rho_{\nu}(\vec{q})\rangle=\frac{1}{n_{0}L L_{s}}\sum_{\kappa\sigma} n^{(s)*}_{\mu-\kappa}h^{(s)}_{\kappa\sigma}(q)n^{*}_{\sigma-\nu}.
\end{equation}

In the following, by exploiting the Ornstein-Zernike equation we will relate the total  correlation function to the direct correlation function in the case of confined liquids. In real space,  the inhomogeneous  Ornstein-Zernike equation for multicomponent liquids between two parallel walls reads
\begin{align}
h_{ij}&(|\vec{r}-\vec{r}'|,z,z')\nonumber\\
=&c_{ij}(|\vec{r}-\vec{r}'|,z,z')+\sum_{k} \int \diff \vec{r}'' \diff z'' c_{ik}(|\vec{r}-\vec{r}'|,z,z'')\nonumber\\
&\times n_{k}(z'')h_{kj}(|\vec{r}''-\vec{r}'|,z,z'),
\end{align}
see Ref.~\cite{Hansen:Theory_of_Simple_Liquids},  where the roman indices $\{i,j,k\}$ refer to different species of a compound liquid. The Ornstein-Zernike defines the direct correlation $c_{ij}(|\vec{r}-\vec{r}'|,z,z')$ in terms of the total correlation function $h_{ij}(|\vec{r}-\vec{r}'|,z,z')$. In our case, $(i,j)\in\{t,h\}$, where $t$ indicates the tracer and $h$ refers to the host liquid species.
Since the tracer particle density is small, $n_{t}(z)=n^{(s)}(z)/N=\mathcal{O}(N^{-1})$, the Ornstein-Zernike equation decouples if one sorts in powers of the particle number $N$. To leading order one obtains the Ornstein-Zernike equation for the solvent,
\begin{align}\label{eq:OZ1}
h&(|\vec{r}-\vec{r}'|,z,z')\nonumber \\
=&c(|\vec{r}-\vec{r}'|,z,z')+ \int \diff \vec{r}'' \diff z'' c(|\vec{r}-\vec{r}''|,z,z'') \nonumber\\
&\times n(z'')h(|\vec{r}''-\vec{r}'|,z'',z'),
\end{align}
where we have suppressed the indices of the host liquid to make contact with our previous notations~\cite{Lang:2012}. The next order yields the mixed  correlation functions, which couple to the spatial positions of the tagged particle,
\begin{align}\label{eq:OZ2}
h^{(s)}&(|\vec{r}-\vec{r}'|,z,z')\\ \nonumber
=&c^{(s)}(|\vec{r}-\vec{r}'|,z,z')+ \int \diff \vec{r}'' \diff z'' c^{(s)}(|\vec{r}-\vec{r}''|,z,z'') \\ \nonumber
&\times n(z'')h(|\vec{r}''-\vec{r}'|,z'',z'),
\end{align}
where we again have employed obvious notation for the mixed correlation functions according to Eq.~\eqref{eq:gs}.

A Fourier decomposition of Eq.~\eqref{eq:OZ1} leads to
\begin{equation}\label{eq:OZblarz}
h_{\mu\nu}(q)=c_{\mu\nu}(q)+\frac{1}{L^2} \sum_{\kappa \lambda} c_{\mu \kappa}(q)n^{*}_{\kappa-\lambda} h_{\lambda\nu}(q).
\end{equation}
Employing the definition of the total correlation function in terms of the van-Hove correlation function $n_{0}G(|\vec{r}-\vec{r}'|,z,z')= n(z) h(|\vec{r}-\vec{r}'|,z,z') n(z')+ n(z) \delta (z-z') \delta (\vec{r}-\vec{r}')$, see Ref.~\cite{Hansen:Theory_of_Simple_Liquids}, the static structure factor $S_{\mu \nu}(q)$~\cite{Lang:2012} can be combined with the total correlation function,
\begin{align}
n_{0}S_{\mu\nu}(q)= \frac{1}{L^2} \sum_{\kappa \lambda}n^{*}_{\mu-\kappa} h_{\kappa\lambda}(q) n^{*}_{\lambda-\nu}+ n^{*}_{\mu-\nu},
\end{align}
and via Eq.~\eqref{eq:OZblarz} with the direct correlation function,
\begin{equation}\label{eq:OZ}
[\bm{S}^{-1}(q)]_{\mu \nu}=\frac{n_{0}}{L^2}[v^*_{\mu-\nu}-c_{\mu \nu}(q)],
\end{equation}
which has been already used to simplify the vertices of the MCT functional for the collective dynamics in confined geometry~\cite{Lang:2012}. Here, the Fourier modes of the local volume $v_{\mu}$ are defined by $\sum_{\kappa}n^*_{\mu-\kappa}v^*_{\kappa-\nu}=L^2\delta_{\mu\nu}$.
Similarly, the mixed correlators  of  the respective Ornstein-Zernike equation, Eq.~\eqref{eq:OZ2}, read
\begin{equation}
h_{\mu\nu}^{(s)}(q)=c_{\mu\nu}^{(s)}(q)+\frac{1}{L^2} \sum_{\kappa \lambda} c^{(s)}_{\mu \kappa}(q)n^*_{\kappa-\lambda} h_{\lambda\nu}(q).
\end{equation}
Eliminating the total correlation function of the liquid in favor  of the structure factor, one obtains the Ornstein-Zernike equation for the tagged particle,
\begin{equation}
h_{\mu\nu}^{(s)}(q)=\frac{n_{0}}{L^2} \sum_{\kappa \lambda}  c^{(s)}_{\mu \kappa} (q)S_{\kappa \lambda}(q) v^*_{\lambda-\nu}.
\end{equation}
The Ornstein-Zernike equation allows us to relate the overlap of the tagged-particle density to the collective density $H^{(s)}_{\mu \nu}(q)=\langle \rho_{\mu}^{(s)}(\vec{q})|\rho_{\nu}(\vec{q})\rangle$ by the direct correlation function and the structure factor of the embedding liquid,
\begin{align}\label{eq:OZ_tagged}
H^{(s)}_{\mu \nu}(q)&=\frac{1}{LL_{s}} \sum_{\kappa \lambda} n_{\mu-\kappa}^{(s)*}c^{(s)}_{\kappa\lambda}(q)S_{\lambda \nu}(q),\nonumber\\
&\equiv \frac{n_0}{LL_{s}} \sum_{\kappa \lambda} S_{\mu\kappa}^{(s)}c^{(s)}_{\kappa\lambda}(q)S_{\lambda \nu}(q).
\end{align}

\section{Operator identity to extract the in-plane dynamics}
\label{sec:Resolvent}

In this appendix we derive the operator identity,
\begin{align}\label{eq:op}
&\mathcal{P}\hat{\bm{S}}^{(s)}(q,z)\mathcal{P}= - \bigl \{z\bm{1}+\mathcal{P}\hat{\bm{K}}^{(s)}(q,z) \nonumber \\
&-\mathcal{P}\hat{\bm{K}}^{(s)}(q,z)\left[ z\bm{1}+\mathcal{Q}\hat{\bm{K}}^{(s)}(q,z)\right]^{-1}\mathcal{Q} \hat{\bm{K}}^{(s)}(q,z)\bigl \}^{-1} \mathcal{P},
\end{align}
where $\mathcal{Q}=[\bm{S}^{(s)}]^{-1}-\mathcal{P}$ and $[\mathcal{P}]_{\mu \nu}=\delta_{\mu 0}\delta_{\nu 0}$, which readily yields the generalized diffusion equation for the in-plane dynamics [see Eq.~\eqref{eq:Diffusion}].

The resolvent calculus technique does not apply directly to Eq.~\eqref{eq:MZ2}, since it differs from the standard resolvent equation $[z\bm{1}-\mathcal{L}]\mathcal{R}(z)=-\bm{1}$ by  the normalization $\bm{S}^{(s)}$.
Therefore, the strategy is to make a detour  performing a basis transformation from the nonorthogonal basis of the density modes $\{|\rho_{\mu}^{(s)}(\vec{q}) \rangle\}$ with $\mu \in \mathbb{Z}$ and $\langle \rho_{\mu}^{(s)}(\vec{q})| \rho_{\nu}^{(s)}(\vec{q})\rangle=S^{(s)}_{\mu\nu}$ to an orthonormalized basis $\{|\Pi_{\mu}^{(s)}(\vec{q})\rangle \}$, i.e., $\langle \Pi_{\mu}^{(s)}(\vec{q})|\Pi_{\nu}^{(s)}(\vec{q}) \rangle= \delta_{\mu \nu}$ for all wave numbers $\vec{q}$. We follow a standard Gram-Schmidt process  adopting the convention,
\begin{align}\label{eq:basis}
& |\Pi^{(s)}_{0}(\vec{q}) \rangle= |\rho^{(s)}_{0}(\vec{q}) \rangle/ \sqrt{ S^{(s)}_{00} }, \nonumber \\
&\dots \nonumber\\
& |\Pi^{(s)}_{\lambda}(\vec{q}) \rangle=\frac{ |\rho^{(s)}_{\lambda}(\vec{q})\rangle-\sum_{\kappa=0}^{\lambda-1}|\Pi^{(s)}_{\kappa}(\vec{q})\rangle \langle \Pi^{(s)}_{\kappa}(\vec{q}) |\rho^{(s)}_{\lambda}(\vec{q})\rangle}{ \norm{\cdot}_{\lambda}},
\end{align}
where the norm $\norm{\cdot}_{\lambda}$ guarantees the normalization of the vector $ |\Pi^{(s)}_{\lambda}(\vec{q}) \rangle$. Since  $|\Pi^{(s)}_{\lambda}(\vec{q}) \rangle$ is a linear combination of the original vectors $|\rho^{(s)}_{\kappa}(\vec{q})\rangle$ with $\kappa=0,\dots,\lambda$ we define a triangular transformation matrix $[\bm{L}]_{\mu\nu}=L_{\mu\nu}$ by
\begin{align}
& | \Pi^{(s)}_{\lambda}(\vec{q}) \rangle = \sum_{\kappa}  | \rho_{\kappa}^{(s)}(\vec{q}) \rangle L_{\kappa \lambda}.
\end{align}
Since the new basis is constructed to be orthonormal one infers
\begin{align}
&\langle \Pi^{(s)}_{\mu}(\vec{q})| \Pi^{(s)}_{\nu}(\vec{q}) \rangle= \sum_{\lambda \kappa} L^*_{\lambda \mu} \langle \rho^{(s)}_{\lambda}(\vec{q}) |\rho^{(s)}_{\kappa}(\vec{q})  \rangle L_{\kappa \nu} \nonumber \\
&=\sum_{\lambda \kappa} L_{\lambda \mu}^* S^{(s)}_{\lambda \kappa}L_{\kappa \nu}=(\bm{L}^{\dagger} \bm{S}^{(s)} \bm{L} )_{\mu \nu}=\delta_{\mu \nu},
\end{align}
or in matrix notation,
\begin{align}\label{eq:idtrans}
 \bm{L}^{\dagger}\bm{S}^{(s)}\bm{L}=\bm{1}.
\end{align}

The equation of motion in the Laplace domain [see Eq.~\eqref{eq:MZ2}], becomes free of a normalization,
\begin{equation}\label{eq:resolvent}
 [z \bm{1}+\tilde{\bm{K}}^{(s)}(q,z) ] \tilde{\bm{S}}^{(s)}(q,z)=- \bm{1},
\end{equation}
where the convention $\tilde{\bm{A}}=\bm{L}^{\dagger} \bm{A}\bm{L}$ is adopted. The indicator $\hat{}$ for the Laplace transformed correlation functions is omitted for a compact notation. Equation~\eqref{eq:resolvent} has the form of a resolvent equation $[z\bm{1}-\mathcal{L}]\mathcal{R}(z)=-\bm{1}$.

The projector onto the distinguished subspace $|\Pi^{(s)}_0(\vec{q})\rangle$ shall be referred to as $\mathcal{P}_{0}$ and its orthogonal complement projecting onto the perpendicular subspace as $\mathcal{Q}_{0}=\bm{1}- \mathcal{P}_{0}$. We suppress the arguments $(q,z)$ for the successive manipulations, which can be found in a similar context in Ref.~\cite{Goetze:Complex_Dynamics}. The starting point is the decomposition,
\begin{align}\label{eq:start}
[z \bm{1} +\tilde{\bm{K}}^{(s)}]\mathcal{P}_{0} \tilde{\bm{S}}^{(s)}+[z \bm{1} +\tilde{\bm{K}}^{(s)} ]\mathcal{Q}_{0} \tilde{\bm{S}}^{(s)}=-\bm{1}.
\end{align}

Sandwiching Eq.~\eqref{eq:start} with $\bm{L} \mathcal{P}_{0}(\dots)\mathcal{P}_{0} \bm{L}^\dagger$ one derives the first identity
\begin{align}\label{eq:id1}
& [z\bm{1} +\mathcal{P}\bm{K}^{(s)}] \mathcal{P} \bm{S}^{(s)}\mathcal{P}+ \mathcal{P} \bm{K}^{(s)}\mathcal{Q} \bm{S}^{(s)}\mathcal{P}= -\mathcal{P},
\end{align}
where  we have defined the pair of generalized projectors
 $\mathcal{P}:=\bm{L}\mathcal{P}_{0}\bm{L}^{\dagger}$ and $\mathcal{Q}:=\bm{L}\mathcal{Q}_{0}\bm{L}^{\dagger}=[\bm{S}^{(s)}]^{-1}-\mathcal{P}$. Note that by construction $\mathcal{P} = |\rho_0^{(s)}(\vec{q}) \rangle [1/S_{00}^{(s)}] \langle \rho_0^{(s)}(\vec{q})|$ is the standard projection operator onto the subspace spanned by $|\rho^{(s)}_{0}(\vec{q}) \rangle$.

 Similarly, a second identity is obtained by sandwiching Eq.~\eqref{eq:start} with $\bm{L} \mathcal{Q}_{0}(\dots)\mathcal{P}_{0} \bm{L}^\dagger$, which results in  $\mathcal{Q} \bm{K}^{(s)}\mathcal{P} \bm{S}^{(s)}\mathcal{P}
+[z \bm{1}+\mathcal{Q} \bm{K}^{(s)}  ]\mathcal{Q} \bm{S}^{(s)}\mathcal{P} =0$. Multiplying from the left by $-\left[z \bm{1}+\mathcal{Q}  \bm{K}^{(s)}  \right]^{-1}$  yields the second identity
\begin{align}\label{eq:id2}
\mathcal{Q}  \bm{S}^{(s)}\mathcal{P} =-\left[ z \bm{1}+\mathcal{Q} \bm{K}^{(s)} \right]^{-1}\mathcal{Q} \bm{K}^{(s)}\mathcal{P} \bm{S}^{(s)}\mathcal{P}.
\end{align}
Inserting Eq.~\eqref{eq:id2} into Eq.~\eqref{eq:id1} one obtains the desired operator identity,
\begin{align}\label{eq:ready}
&\mathcal{P}\hat{\bm{S}}^{(s)}(q,z)\mathcal{P}= - \bigl \{z\bm{1}+\mathcal{P}\hat{\bm{K}}^{(s)}(q,z) \nonumber \\
&-\mathcal{P}\hat{\bm{K}}^{(s)}(q,z)\left[ z\bm{1}+\mathcal{Q}\hat{\bm{K}}^{(s)}(q,z)\right]^{-1}\mathcal{Q} \hat{\bm{K}}^{(s)}(q,z)\bigl \}^{-1} \mathcal{P},
\end{align}
where the arguments and the symbol $\hat{}$ to indicate the Laplace transformed quantities have been reintroduced.   Let us emphasize that  the
 operator identity, Eq.~\eqref{eq:ready}, is covariant, i.e., independent of the rendered transformation $\bm{L}$ and hence applicable on general grounds.


\begin{acknowledgments}
We are indebted to Rolf Schilling for valuable comments and careful reading of the manuscript.
This work has been supported by the Deutsche Forschungsgemeinschaft DFG via the  Research Unit FOR1394 ``Nonlinear Response to Probe Vitrification''.
S.L. is thankful for support from the Cluster of Excellence ``Engineering of Advanced Materials'' at the
University of Erlangen-Nuremberg.
\end{acknowledgments}

\bibliographystyle{apsrev4-1}


\begin{thebibliography}{74}%
\makeatletter
\providecommand \@ifxundefined [1]{%
 \@ifx{#1\undefined}
}%
\providecommand \@ifnum [1]{%
 \ifnum #1\expandafter \@firstoftwo
 \else \expandafter \@secondoftwo
 \fi
}%
\providecommand \@ifx [1]{%
 \ifx #1\expandafter \@firstoftwo
 \else \expandafter \@secondoftwo
 \fi
}%
\providecommand \natexlab [1]{#1}%
\providecommand \enquote  [1]{``#1''}%
\providecommand \bibnamefont  [1]{#1}%
\providecommand \bibfnamefont [1]{#1}%
\providecommand \citenamefont [1]{#1}%
\providecommand \href@noop [0]{\@secondoftwo}%
\providecommand \href [0]{\begingroup \@sanitize@url \@href}%
\providecommand \@href[1]{\@@startlink{#1}\@@href}%
\providecommand \@@href[1]{\endgroup#1\@@endlink}%
\providecommand \@sanitize@url [0]{\catcode `\\12\catcode `\$12\catcode
  `\&12\catcode `\#12\catcode `\^12\catcode `\_12\catcode `\%12\relax}%
\providecommand \@@startlink[1]{}%
\providecommand \@@endlink[0]{}%
\providecommand \url  [0]{\begingroup\@sanitize@url \@url }%
\providecommand \@url [1]{\endgroup\@href {#1}{\urlprefix }}%
\providecommand \urlprefix  [0]{URL }%
\providecommand \Eprint [0]{\href }%
\providecommand \doibase [0]{http://dx.doi.org/}%
\providecommand \selectlanguage [0]{\@gobble}%
\providecommand \bibinfo  [0]{\@secondoftwo}%
\providecommand \bibfield  [0]{\@secondoftwo}%
\providecommand \translation [1]{[#1]}%
\providecommand \BibitemOpen [0]{}%
\providecommand \bibitemStop [0]{}%
\providecommand \bibitemNoStop [0]{.\EOS\space}%
\providecommand \EOS [0]{\spacefactor3000\relax}%
\providecommand \BibitemShut  [1]{\csname bibitem#1\endcsname}%
\let\auto@bib@innerbib\@empty
\bibitem [{\citenamefont {G\"otze}(2009)}]{Goetze:Complex_Dynamics}%
  \BibitemOpen
  \bibfield  {author} {\bibinfo {author} {\bibfnamefont {W.}~\bibnamefont
  {G\"otze}},\ }\href@noop {} {\emph {\bibinfo {title} {Complex Dynamics of
  Glass-Forming Liquids -- A Mode-Coupling Theory}}}\ (\bibinfo  {publisher}
  {Oxford},\ \bibinfo {address} {Oxford},\ \bibinfo {year} {2009})\BibitemShut
  {NoStop}%
\bibitem [{\citenamefont {Bengtzelius}\ \emph {et~al.}(1984)\citenamefont
  {Bengtzelius}, \citenamefont {G\"otze},\ and\ \citenamefont
  {Sj\"olander}}]{Bengtzelius:1984}%
  \BibitemOpen
  \bibfield  {author} {\bibinfo {author} {\bibfnamefont {U.}~\bibnamefont
  {Bengtzelius}}, \bibinfo {author} {\bibfnamefont {W.}~\bibnamefont
  {G\"otze}}, \ and\ \bibinfo {author} {\bibfnamefont {A.}~\bibnamefont
  {Sj\"olander}},\ }\href {http://stacks.iop.org/0022-3719/17/i=33/a=005}
  {\bibfield  {journal} {\bibinfo  {journal} {J. Phys. C: Solid State Phys.}\
  }\textbf {\bibinfo {volume} {17}},\ \bibinfo {pages} {5915} (\bibinfo {year}
  {1984})}\BibitemShut {NoStop}%
\bibitem [{\citenamefont {Li}\ \emph {et~al.}(1992)\citenamefont {Li},
  \citenamefont {Du}, \citenamefont {Chen}, \citenamefont {Cummins},\ and\
  \citenamefont {Tao}}]{Li:1992}%
  \BibitemOpen
  \bibfield  {author} {\bibinfo {author} {\bibfnamefont {G.}~\bibnamefont
  {Li}}, \bibinfo {author} {\bibfnamefont {W.~M.}\ \bibnamefont {Du}}, \bibinfo
  {author} {\bibfnamefont {X.~K.}\ \bibnamefont {Chen}}, \bibinfo {author}
  {\bibfnamefont {H.~Z.}\ \bibnamefont {Cummins}}, \ and\ \bibinfo {author}
  {\bibfnamefont {N.~J.}\ \bibnamefont {Tao}},\ }\href {\doibase
  10.1103/PhysRevA.45.3867} {\bibfield  {journal} {\bibinfo  {journal} {Phys.
  Rev. A}\ }\textbf {\bibinfo {volume} {45}},\ \bibinfo {pages} {3867}
  (\bibinfo {year} {1992})}\BibitemShut {NoStop}%
\bibitem [{\citenamefont {van Megen}\ and\ \citenamefont
  {Underwood}(1993)}]{Megen:1993a}%
  \BibitemOpen
  \bibfield  {author} {\bibinfo {author} {\bibfnamefont {W.}~\bibnamefont {van
  Megen}}\ and\ \bibinfo {author} {\bibfnamefont {S.~M.}\ \bibnamefont
  {Underwood}},\ }\href {\doibase 10.1103/PhysRevLett.70.2766} {\bibfield
  {journal} {\bibinfo  {journal} {Phys. Rev. Lett.}\ }\textbf {\bibinfo
  {volume} {70}},\ \bibinfo {pages} {2766} (\bibinfo {year}
  {1993})}\BibitemShut {NoStop}%
\bibitem [{\citenamefont {Wuttke}\ \emph {et~al.}(1994)\citenamefont {Wuttke},
  \citenamefont {Hernandez}, \citenamefont {Li}, \citenamefont {Coddens},
  \citenamefont {Cummins}, \citenamefont {Fujara}, \citenamefont {Petry},\ and\
  \citenamefont {Sillescu}}]{Wuttke:1994}%
  \BibitemOpen
  \bibfield  {author} {\bibinfo {author} {\bibfnamefont {J.}~\bibnamefont
  {Wuttke}}, \bibinfo {author} {\bibfnamefont {J.}~\bibnamefont {Hernandez}},
  \bibinfo {author} {\bibfnamefont {G.}~\bibnamefont {Li}}, \bibinfo {author}
  {\bibfnamefont {G.}~\bibnamefont {Coddens}}, \bibinfo {author} {\bibfnamefont
  {H.~Z.}\ \bibnamefont {Cummins}}, \bibinfo {author} {\bibfnamefont
  {F.}~\bibnamefont {Fujara}}, \bibinfo {author} {\bibfnamefont
  {W.}~\bibnamefont {Petry}}, \ and\ \bibinfo {author} {\bibfnamefont
  {H.}~\bibnamefont {Sillescu}},\ }\href {\doibase 10.1103/PhysRevLett.72.3052}
  {\bibfield  {journal} {\bibinfo  {journal} {Phys. Rev. Lett.}\ }\textbf
  {\bibinfo {volume} {72}},\ \bibinfo {pages} {3052} (\bibinfo {year}
  {1994})}\BibitemShut {NoStop}%
\bibitem [{\citenamefont {Torre}\ \emph {et~al.}(2000)\citenamefont {Torre},
  \citenamefont {Bartolini}, \citenamefont {Ricci},\ and\ \citenamefont
  {Pick}}]{Torre:2000}%
  \BibitemOpen
  \bibfield  {author} {\bibinfo {author} {\bibfnamefont {R.}~\bibnamefont
  {Torre}}, \bibinfo {author} {\bibfnamefont {P.}~\bibnamefont {Bartolini}},
  \bibinfo {author} {\bibfnamefont {M.}~\bibnamefont {Ricci}}, \ and\ \bibinfo
  {author} {\bibfnamefont {R.~M.}\ \bibnamefont {Pick}},\ }\href
  {http://stacks.iop.org/0295-5075/52/i=3/a=324} {\bibfield  {journal}
  {\bibinfo  {journal} {EPL (Europhysics Letters)}\ }\textbf {\bibinfo {volume}
  {52}},\ \bibinfo {pages} {324} (\bibinfo {year} {2000})}\BibitemShut
  {NoStop}%
\bibitem [{\citenamefont {van Megen}\ \emph {et~al.}(1998)\citenamefont {van
  Megen}, \citenamefont {Mortensen}, \citenamefont {Williams},\ and\
  \citenamefont {M\"uller}}]{Megen:1998}%
  \BibitemOpen
  \bibfield  {author} {\bibinfo {author} {\bibfnamefont {W.}~\bibnamefont {van
  Megen}}, \bibinfo {author} {\bibfnamefont {T.~C.}\ \bibnamefont {Mortensen}},
  \bibinfo {author} {\bibfnamefont {S.~R.}\ \bibnamefont {Williams}}, \ and\
  \bibinfo {author} {\bibfnamefont {J.}~\bibnamefont {M\"uller}},\ }\href
  {\doibase 10.1103/PhysRevE.58.6073} {\bibfield  {journal} {\bibinfo
  {journal} {Phys. Rev. E}\ }\textbf {\bibinfo {volume} {58}},\ \bibinfo
  {pages} {6073} (\bibinfo {year} {1998})}\BibitemShut {NoStop}%
\bibitem [{\citenamefont {Singh}\ \emph {et~al.}(1998)\citenamefont {Singh},
  \citenamefont {Li}, \citenamefont {G\"otze}, \citenamefont {Fuchs},
  \citenamefont {Franosch},\ and\ \citenamefont {Cummins}}]{Singh:1998}%
  \BibitemOpen
  \bibfield  {author} {\bibinfo {author} {\bibfnamefont {A.~P.}\ \bibnamefont
  {Singh}}, \bibinfo {author} {\bibfnamefont {G.}~\bibnamefont {Li}}, \bibinfo
  {author} {\bibfnamefont {W.}~\bibnamefont {G\"otze}}, \bibinfo {author}
  {\bibfnamefont {M.}~\bibnamefont {Fuchs}}, \bibinfo {author} {\bibfnamefont
  {T.}~\bibnamefont {Franosch}}, \ and\ \bibinfo {author} {\bibfnamefont
  {H.~Z.}\ \bibnamefont {Cummins}},\ }\href {\doibase
  10.1016/S0022-3093(98)00583-3} {\bibfield  {journal} {\bibinfo  {journal} {J.
  Non-Cryst. Solids}\ }\textbf {\bibinfo {volume} {235-237}},\ \bibinfo {pages}
  {66 } (\bibinfo {year} {1998})}\BibitemShut {NoStop}%
\bibitem [{\citenamefont {G\"otze}(1999)}]{Goetze:1999}%
  \BibitemOpen
  \bibfield  {author} {\bibinfo {author} {\bibfnamefont {W.}~\bibnamefont
  {G\"otze}},\ }\href {http://stacks.iop.org/0953-8984/11/i=10A/a=002}
  {\bibfield  {journal} {\bibinfo  {journal} {J. Phys.: Condens. Matter}\
  }\textbf {\bibinfo {volume} {11}},\ \bibinfo {pages} {A1} (\bibinfo {year}
  {1999})}\BibitemShut {NoStop}%
\bibitem [{\citenamefont {Kob}\ and\ \citenamefont
  {Andersen}(1994)}]{Kob:1994}%
  \BibitemOpen
  \bibfield  {author} {\bibinfo {author} {\bibfnamefont {W.}~\bibnamefont
  {Kob}}\ and\ \bibinfo {author} {\bibfnamefont {H.~C.}\ \bibnamefont
  {Andersen}},\ }\href {\doibase 10.1103/PhysRevLett.73.1376} {\bibfield
  {journal} {\bibinfo  {journal} {Phys. Rev. Lett.}\ }\textbf {\bibinfo
  {volume} {73}},\ \bibinfo {pages} {1376} (\bibinfo {year}
  {1994})}\BibitemShut {NoStop}%
\bibitem [{\citenamefont {Kob}\ and\ \citenamefont
  {Andersen}(1995{\natexlab{a}})}]{Kob:1995a}%
  \BibitemOpen
  \bibfield  {author} {\bibinfo {author} {\bibfnamefont {W.}~\bibnamefont
  {Kob}}\ and\ \bibinfo {author} {\bibfnamefont {H.~C.}\ \bibnamefont
  {Andersen}},\ }\href {\doibase 10.1103/PhysRevE.51.4626} {\bibfield
  {journal} {\bibinfo  {journal} {Phys. Rev. E}\ }\textbf {\bibinfo {volume}
  {51}},\ \bibinfo {pages} {4626} (\bibinfo {year}
  {1995}{\natexlab{a}})}\BibitemShut {NoStop}%
\bibitem [{\citenamefont {Kob}\ and\ \citenamefont
  {Andersen}(1995{\natexlab{b}})}]{Kob:1995b}%
  \BibitemOpen
  \bibfield  {author} {\bibinfo {author} {\bibfnamefont {W.}~\bibnamefont
  {Kob}}\ and\ \bibinfo {author} {\bibfnamefont {H.~C.}\ \bibnamefont
  {Andersen}},\ }\href {\doibase {10.1103/PhysRevE.52.4134}} {\bibfield
  {journal} {\bibinfo  {journal} {{Phys. Rev. E}}\ }\textbf {\bibinfo {volume}
  {{52}}},\ \bibinfo {pages} {4134} (\bibinfo {year}
  {{1995}}{\natexlab{b}})}\BibitemShut {NoStop}%
\bibitem [{\citenamefont {K\"ammerer}\ \emph
  {et~al.}(1998{\natexlab{a}})\citenamefont {K\"ammerer}, \citenamefont {Kob},\
  and\ \citenamefont {Schilling}}]{Kaemmerer:1998}%
  \BibitemOpen
  \bibfield  {author} {\bibinfo {author} {\bibfnamefont {S.}~\bibnamefont
  {K\"ammerer}}, \bibinfo {author} {\bibfnamefont {W.}~\bibnamefont {Kob}}, \
  and\ \bibinfo {author} {\bibfnamefont {R.}~\bibnamefont {Schilling}},\ }\href
  {\doibase 10.1103/PhysRevE.58.2131} {\bibfield  {journal} {\bibinfo
  {journal} {Phys. Rev. E}\ }\textbf {\bibinfo {volume} {58}},\ \bibinfo
  {pages} {2131} (\bibinfo {year} {1998}{\natexlab{a}})}\BibitemShut {NoStop}%
\bibitem [{\citenamefont {K\"ammerer}\ \emph
  {et~al.}(1998{\natexlab{b}})\citenamefont {K\"ammerer}, \citenamefont {Kob},\
  and\ \citenamefont {Schilling}}]{Kaemmerer:1998a}%
  \BibitemOpen
  \bibfield  {author} {\bibinfo {author} {\bibfnamefont {S.}~\bibnamefont
  {K\"ammerer}}, \bibinfo {author} {\bibfnamefont {W.}~\bibnamefont {Kob}}, \
  and\ \bibinfo {author} {\bibfnamefont {R.}~\bibnamefont {Schilling}},\ }\href
  {\doibase 10.1103/PhysRevE.58.2141} {\bibfield  {journal} {\bibinfo
  {journal} {Phys. Rev. E}\ }\textbf {\bibinfo {volume} {58}},\ \bibinfo
  {pages} {2141} (\bibinfo {year} {1998}{\natexlab{b}})}\BibitemShut {NoStop}%
\bibitem [{\citenamefont {Franosch}\ \emph {et~al.}(1997)\citenamefont
  {Franosch}, \citenamefont {Fuchs}, \citenamefont {G{\"o}tze}, \citenamefont
  {Mayr},\ and\ \citenamefont {Singh}}]{Franosch:1997}%
  \BibitemOpen
  \bibfield  {author} {\bibinfo {author} {\bibfnamefont {T.}~\bibnamefont
  {Franosch}}, \bibinfo {author} {\bibfnamefont {M.}~\bibnamefont {Fuchs}},
  \bibinfo {author} {\bibfnamefont {W.}~\bibnamefont {G{\"o}tze}}, \bibinfo
  {author} {\bibfnamefont {M.~R.}\ \bibnamefont {Mayr}}, \ and\ \bibinfo
  {author} {\bibfnamefont {A.~P.}\ \bibnamefont {Singh}},\ }\href {\doibase
  10.1103/PhysRevE.56.5659} {\bibfield  {journal} {\bibinfo  {journal} {{Phys.
  Rev. E}}\ }\textbf {\bibinfo {volume} {{56}}},\ \bibinfo {pages} {5659}
  (\bibinfo {year} {1997})}\BibitemShut {NoStop}%
\bibitem [{\citenamefont {Fuchs}\ \emph {et~al.}(1998)\citenamefont {Fuchs},
  \citenamefont {G\"otze},\ and\ \citenamefont {Mayr}}]{Fuchs:1998}%
  \BibitemOpen
  \bibfield  {author} {\bibinfo {author} {\bibfnamefont {M.}~\bibnamefont
  {Fuchs}}, \bibinfo {author} {\bibfnamefont {W.}~\bibnamefont {G\"otze}}, \
  and\ \bibinfo {author} {\bibfnamefont {M.~R.}\ \bibnamefont {Mayr}},\ }\href
  {\doibase 10.1103/PhysRevE.58.3384} {\bibfield  {journal} {\bibinfo
  {journal} {Phys. Rev. E}\ }\textbf {\bibinfo {volume} {58}},\ \bibinfo
  {pages} {3384} (\bibinfo {year} {1998})}\BibitemShut {NoStop}%
\bibitem [{\citenamefont {Krakoviack}(2005)}]{Krakoviack:2005}%
  \BibitemOpen
  \bibfield  {author} {\bibinfo {author} {\bibfnamefont {V.}~\bibnamefont
  {Krakoviack}},\ }\href {\doibase 10.1103/PhysRevLett.94.065703} {\bibfield
  {journal} {\bibinfo  {journal} {Phys. Rev. Lett.}\ }\textbf {\bibinfo
  {volume} {94}},\ \bibinfo {pages} {065703} (\bibinfo {year}
  {2005})}\BibitemShut {NoStop}%
\bibitem [{\citenamefont {Krakoviack}(2007)}]{Krakoviack:2007}%
  \BibitemOpen
  \bibfield  {author} {\bibinfo {author} {\bibfnamefont {V.}~\bibnamefont
  {Krakoviack}},\ }\href {\doibase 10.1103/PhysRevE.75.031503} {\bibfield
  {journal} {\bibinfo  {journal} {Phys. Rev. E}\ }\textbf {\bibinfo {volume}
  {75}},\ \bibinfo {eid} {031503} (\bibinfo {year} {2007})}\BibitemShut
  {NoStop}%
\bibitem [{\citenamefont {Krakoviack}(2011)}]{Krakoviack:2011}%
  \BibitemOpen
  \bibfield  {author} {\bibinfo {author} {\bibfnamefont {V.}~\bibnamefont
  {Krakoviack}},\ }\href {\doibase 10.1103/PhysRevE.84.050501} {\bibfield
  {journal} {\bibinfo  {journal} {Phys. Rev. E}\ }\textbf {\bibinfo {volume}
  {84}},\ \bibinfo {pages} {050501} (\bibinfo {year} {2011})}\BibitemShut
  {NoStop}%
\bibitem [{\citenamefont {Kurzidim}\ \emph {et~al.}(2009)\citenamefont
  {Kurzidim}, \citenamefont {Coslovich},\ and\ \citenamefont
  {Kahl}}]{Kurzidim:2009}%
  \BibitemOpen
  \bibfield  {author} {\bibinfo {author} {\bibfnamefont {J.}~\bibnamefont
  {Kurzidim}}, \bibinfo {author} {\bibfnamefont {D.}~\bibnamefont {Coslovich}},
  \ and\ \bibinfo {author} {\bibfnamefont {G.}~\bibnamefont {Kahl}},\ }\href
  {\doibase 10.1103/PhysRevLett.103.138303} {\bibfield  {journal} {\bibinfo
  {journal} {Phys. Rev. Lett.}\ }\textbf {\bibinfo {volume} {103}},\ \bibinfo
  {pages} {138303} (\bibinfo {year} {2009})}\BibitemShut {NoStop}%
\bibitem [{\citenamefont {Kim}\ \emph {et~al.}(2009)\citenamefont {Kim},
  \citenamefont {Miyazaki},\ and\ \citenamefont {Saito}}]{Kim:2009}%
  \BibitemOpen
  \bibfield  {author} {\bibinfo {author} {\bibfnamefont {K.}~\bibnamefont
  {Kim}}, \bibinfo {author} {\bibfnamefont {K.}~\bibnamefont {Miyazaki}}, \
  and\ \bibinfo {author} {\bibfnamefont {S.}~\bibnamefont {Saito}},\ }\href
  {http://stacks.iop.org/0295-5075/88/i=3/a=36002} {\bibfield  {journal}
  {\bibinfo  {journal} {EPL}\ }\textbf {\bibinfo {volume} {88}},\ \bibinfo
  {pages} {36002} (\bibinfo {year} {2009})}\BibitemShut {NoStop}%
\bibitem [{\citenamefont {Horbach}\ \emph {et~al.}(2002)\citenamefont
  {Horbach}, \citenamefont {Kob},\ and\ \citenamefont {Binder}}]{Horbach:2002}%
  \BibitemOpen
  \bibfield  {author} {\bibinfo {author} {\bibfnamefont {J.}~\bibnamefont
  {Horbach}}, \bibinfo {author} {\bibfnamefont {W.}~\bibnamefont {Kob}}, \ and\
  \bibinfo {author} {\bibfnamefont {K.}~\bibnamefont {Binder}},\ }\href
  {\doibase 10.1103/PhysRevLett.88.125502} {\bibfield  {journal} {\bibinfo
  {journal} {Phys. Rev. Lett.}\ }\textbf {\bibinfo {volume} {88}},\ \bibinfo
  {pages} {125502} (\bibinfo {year} {2002})}\BibitemShut {NoStop}%
\bibitem [{\citenamefont {Voigtmann}\ and\ \citenamefont
  {Horbach}(2006)}]{Voigtmann:2006}%
  \BibitemOpen
  \bibfield  {author} {\bibinfo {author} {\bibfnamefont {{\relax
  Th}.}~\bibnamefont {Voigtmann}}\ and\ \bibinfo {author} {\bibfnamefont
  {J.}~\bibnamefont {Horbach}},\ }\href
  {http://stacks.iop.org/0295-5075/74/i=3/a=459} {\bibfield  {journal}
  {\bibinfo  {journal} {EPL}\ }\textbf {\bibinfo {volume} {74}},\ \bibinfo
  {pages} {459} (\bibinfo {year} {2006})}\BibitemShut {NoStop}%
\bibitem [{\citenamefont {Szamel}\ and\ \citenamefont
  {Flenner}(2013)}]{Szamel:2013}%
  \BibitemOpen
  \bibfield  {author} {\bibinfo {author} {\bibfnamefont {G.}~\bibnamefont
  {Szamel}}\ and\ \bibinfo {author} {\bibfnamefont {E.}~\bibnamefont
  {Flenner}},\ }\href {http://stacks.iop.org/0295-5075/101/i=6/a=66005}
  {\bibfield  {journal} {\bibinfo  {journal} {EPL}\ }\textbf {\bibinfo {volume}
  {101}},\ \bibinfo {pages} {66005} (\bibinfo {year} {2013})}\BibitemShut
  {NoStop}%
\bibitem [{\citenamefont {Fehr}\ and\ \citenamefont
  {L\"owen}(1995)}]{Fehr:1995}%
  \BibitemOpen
  \bibfield  {author} {\bibinfo {author} {\bibfnamefont {T.}~\bibnamefont
  {Fehr}}\ and\ \bibinfo {author} {\bibfnamefont {H.}~\bibnamefont {L\"owen}},\
  }\href {\doibase 10.1103/PhysRevE.52.4016} {\bibfield  {journal} {\bibinfo
  {journal} {Phys. Rev. E}\ }\textbf {\bibinfo {volume} {52}},\ \bibinfo
  {pages} {4016} (\bibinfo {year} {1995})}\BibitemShut {NoStop}%
\bibitem [{\citenamefont {Scheidler}\ \emph
  {et~al.}(2000{\natexlab{a}})\citenamefont {Scheidler}, \citenamefont {Kob},\
  and\ \citenamefont {Binder}}]{Scheidler:2000b}%
  \BibitemOpen
  \bibfield  {author} {\bibinfo {author} {\bibfnamefont {P.}~\bibnamefont
  {Scheidler}}, \bibinfo {author} {\bibfnamefont {W.}~\bibnamefont {Kob}}, \
  and\ \bibinfo {author} {\bibfnamefont {K.}~\bibnamefont {Binder}},\ }\href
  {http://stacks.iop.org/0295-5075/52/i=3/a=277} {\bibfield  {journal}
  {\bibinfo  {journal} {EPL}\ }\textbf {\bibinfo {volume} {52}},\ \bibinfo
  {pages} {277} (\bibinfo {year} {2000}{\natexlab{a}})}\BibitemShut {NoStop}%
\bibitem [{\citenamefont {Scheidler}\ \emph
  {et~al.}(2000{\natexlab{b}})\citenamefont {Scheidler}, \citenamefont {Kob},\
  and\ \citenamefont {Binder}}]{Scheidler:2000a}%
  \BibitemOpen
  \bibfield  {author} {\bibinfo {author} {\bibfnamefont {P.}~\bibnamefont
  {Scheidler}}, \bibinfo {author} {\bibfnamefont {W.}~\bibnamefont {Kob}}, \
  and\ \bibinfo {author} {\bibfnamefont {K.}~\bibnamefont {Binder}},\ }\href
  {\doibase 10.1051/jp4:2000706} {\bibfield  {journal} {\bibinfo  {journal} {J.
  Phys. IV France}\ }\textbf {\bibinfo {volume} {10}},\ \bibinfo {pages} {Pr7}
  (\bibinfo {year} {2000}{\natexlab{b}})}\BibitemShut {NoStop}%
\bibitem [{\citenamefont {Scheidler}\ \emph {et~al.}(2002)\citenamefont
  {Scheidler}, \citenamefont {Kob},\ and\ \citenamefont
  {Binder}}]{Scheidler:2002}%
  \BibitemOpen
  \bibfield  {author} {\bibinfo {author} {\bibfnamefont {P.}~\bibnamefont
  {Scheidler}}, \bibinfo {author} {\bibfnamefont {W.}~\bibnamefont {Kob}}, \
  and\ \bibinfo {author} {\bibfnamefont {K.}~\bibnamefont {Binder}},\ }\href
  {http://stacks.iop.org/0295-5075/59/i=5/a=701} {\bibfield  {journal}
  {\bibinfo  {journal} {EPL}\ }\textbf {\bibinfo {volume} {59}},\ \bibinfo
  {pages} {701} (\bibinfo {year} {2002})}\BibitemShut {NoStop}%
\bibitem [{\citenamefont {Scheidler}\ \emph {et~al.}(2004)\citenamefont
  {Scheidler}, \citenamefont {Kob},\ and\ \citenamefont
  {Binder}}]{Scheidler:2004}%
  \BibitemOpen
  \bibfield  {author} {\bibinfo {author} {\bibfnamefont {P.}~\bibnamefont
  {Scheidler}}, \bibinfo {author} {\bibfnamefont {W.}~\bibnamefont {Kob}}, \
  and\ \bibinfo {author} {\bibfnamefont {K.}~\bibnamefont {Binder}},\ }\href
  {\doibase 10.1021/jp036593s} {\bibfield  {journal} {\bibinfo  {journal} {J.
  Phys. Chem. B}\ }\textbf {\bibinfo {volume} {108}},\ \bibinfo {pages} {6673}
  (\bibinfo {year} {2004})}\BibitemShut {NoStop}%
\bibitem [{\citenamefont {Teboul}\ and\ \citenamefont {{Alba
  Simionesco}}(2002)}]{Teboul:2002}%
  \BibitemOpen
  \bibfield  {author} {\bibinfo {author} {\bibfnamefont {V.}~\bibnamefont
  {Teboul}}\ and\ \bibinfo {author} {\bibfnamefont {C.}~\bibnamefont {{Alba
  Simionesco}}},\ }\href {\doibase 10.1088/0953-8984/14/23/304} {\bibfield
  {journal} {\bibinfo  {journal} {J. Phys. Condens. Matter}\ }\textbf {\bibinfo
  {volume} {14}},\ \bibinfo {pages} {5699} (\bibinfo {year}
  {2002})}\BibitemShut {NoStop}%
\bibitem [{\citenamefont {Varnik}\ \emph {et~al.}(2002)\citenamefont {Varnik},
  \citenamefont {Baschnagel},\ and\ \citenamefont {Binder}}]{Varnik:2002}%
  \BibitemOpen
  \bibfield  {author} {\bibinfo {author} {\bibfnamefont {F.}~\bibnamefont
  {Varnik}}, \bibinfo {author} {\bibfnamefont {J.}~\bibnamefont {Baschnagel}},
  \ and\ \bibinfo {author} {\bibfnamefont {K.}~\bibnamefont {Binder}},\ }\href
  {\doibase 10.1103/PhysRevE.65.021507} {\bibfield  {journal} {\bibinfo
  {journal} {Phys. Rev. E}\ }\textbf {\bibinfo {volume} {65}},\ \bibinfo
  {pages} {021507} (\bibinfo {year} {2002})}\BibitemShut {NoStop}%
\bibitem [{\citenamefont {Baschnagel}\ and\ \citenamefont
  {Varnik}(2005)}]{Baschnagel:2005}%
  \BibitemOpen
  \bibfield  {author} {\bibinfo {author} {\bibfnamefont {J.}~\bibnamefont
  {Baschnagel}}\ and\ \bibinfo {author} {\bibfnamefont {F.}~\bibnamefont
  {Varnik}},\ }\href {\doibase 10.1088/0953-8984/17/32/R02} {\bibfield
  {journal} {\bibinfo  {journal} {J. Phys. Condens. Matter}\ }\textbf {\bibinfo
  {volume} {17}},\ \bibinfo {pages} {R851} (\bibinfo {year}
  {2005})}\BibitemShut {NoStop}%
\bibitem [{\citenamefont {Nugent}\ \emph {et~al.}(2007)\citenamefont {Nugent},
  \citenamefont {Edmond}, \citenamefont {Patel},\ and\ \citenamefont
  {Weeks}}]{Nugent:2007}%
  \BibitemOpen
  \bibfield  {author} {\bibinfo {author} {\bibfnamefont {C.~R.}\ \bibnamefont
  {Nugent}}, \bibinfo {author} {\bibfnamefont {K.~V.}\ \bibnamefont {Edmond}},
  \bibinfo {author} {\bibfnamefont {H.~N.}\ \bibnamefont {Patel}}, \ and\
  \bibinfo {author} {\bibfnamefont {E.~R.}\ \bibnamefont {Weeks}},\ }\href
  {\doibase 10.1103/PhysRevLett.99.025702} {\bibfield  {journal} {\bibinfo
  {journal} {Phys. Rev. Lett.}\ }\textbf {\bibinfo {volume} {99}},\ \bibinfo
  {pages} {025702} (\bibinfo {year} {2007})}\BibitemShut {NoStop}%
\bibitem [{\citenamefont {Eral}\ \emph {et~al.}(2009)\citenamefont {Eral},
  \citenamefont {van~den Ende}, \citenamefont {Mugele},\ and\ \citenamefont
  {Duits}}]{Eral:2009}%
  \BibitemOpen
  \bibfield  {author} {\bibinfo {author} {\bibfnamefont {H.~B.}\ \bibnamefont
  {Eral}}, \bibinfo {author} {\bibfnamefont {D.}~\bibnamefont {van~den Ende}},
  \bibinfo {author} {\bibfnamefont {F.}~\bibnamefont {Mugele}}, \ and\ \bibinfo
  {author} {\bibfnamefont {M.~H.~G.}\ \bibnamefont {Duits}},\ }\href {\doibase
  10.1103/PhysRevE.80.061403} {\bibfield  {journal} {\bibinfo  {journal} {Phys.
  Rev. E}\ }\textbf {\bibinfo {volume} {80}},\ \bibinfo {pages} {061403}
  (\bibinfo {year} {2009})}\BibitemShut {NoStop}%
\bibitem [{\citenamefont {Eral}\ \emph {et~al.}(2011)\citenamefont {Eral},
  \citenamefont {Mugele},\ and\ \citenamefont {Duits}}]{Eral:2011}%
  \BibitemOpen
  \bibfield  {author} {\bibinfo {author} {\bibfnamefont {H.~B.}\ \bibnamefont
  {Eral}}, \bibinfo {author} {\bibfnamefont {F.}~\bibnamefont {Mugele}}, \ and\
  \bibinfo {author} {\bibfnamefont {M.~H.~G.}\ \bibnamefont {Duits}},\ }\href
  {\doibase 10.1021/la2024764} {\bibfield  {journal} {\bibinfo  {journal}
  {Langmuir}\ }\textbf {\bibinfo {volume} {27}},\ \bibinfo {pages} {12297}
  (\bibinfo {year} {2011})}\BibitemShut {NoStop}%
\bibitem [{\citenamefont {Edmond}\ \emph {et~al.}(2012)\citenamefont {Edmond},
  \citenamefont {Nugent},\ and\ \citenamefont {Weeks}}]{Edmond:2012}%
  \BibitemOpen
  \bibfield  {author} {\bibinfo {author} {\bibfnamefont {K.~V.}\ \bibnamefont
  {Edmond}}, \bibinfo {author} {\bibfnamefont {C.~R.}\ \bibnamefont {Nugent}},
  \ and\ \bibinfo {author} {\bibfnamefont {E.~R.}\ \bibnamefont {Weeks}},\
  }\href {\doibase 10.1103/PhysRevE.85.041401} {\bibfield  {journal} {\bibinfo
  {journal} {Phys. Rev. E}\ }\textbf {\bibinfo {volume} {85}},\ \bibinfo
  {pages} {041401} (\bibinfo {year} {2012})}\BibitemShut {NoStop}%
\bibitem [{\citenamefont {Gallo}\ \emph
  {et~al.}(2000{\natexlab{a}})\citenamefont {Gallo}, \citenamefont {Rovere},\
  and\ \citenamefont {Spohr}}]{Gallo:2000a}%
  \BibitemOpen
  \bibfield  {author} {\bibinfo {author} {\bibfnamefont {P.}~\bibnamefont
  {Gallo}}, \bibinfo {author} {\bibfnamefont {M.}~\bibnamefont {Rovere}}, \
  and\ \bibinfo {author} {\bibfnamefont {E.}~\bibnamefont {Spohr}},\ }\href
  {\doibase 10.1103/PhysRevLett.85.4317} {\bibfield  {journal} {\bibinfo
  {journal} {Phys. Rev. Lett.}\ }\textbf {\bibinfo {volume} {85}},\ \bibinfo
  {pages} {4317} (\bibinfo {year} {2000}{\natexlab{a}})}\BibitemShut {NoStop}%
\bibitem [{\citenamefont {Gallo}\ \emph
  {et~al.}(2000{\natexlab{b}})\citenamefont {Gallo}, \citenamefont {Rovere},\
  and\ \citenamefont {Spohr}}]{Gallo:2000b}%
  \BibitemOpen
  \bibfield  {author} {\bibinfo {author} {\bibfnamefont {P.}~\bibnamefont
  {Gallo}}, \bibinfo {author} {\bibfnamefont {M.}~\bibnamefont {Rovere}}, \
  and\ \bibinfo {author} {\bibfnamefont {E.}~\bibnamefont {Spohr}},\ }\href
  {\doibase 10.1063/1.1328073} {\bibfield  {journal} {\bibinfo  {journal} {J.
  Chem. Phys.}\ }\textbf {\bibinfo {volume} {113}},\ \bibinfo {pages} {11324}
  (\bibinfo {year} {2000}{\natexlab{b}})}\BibitemShut {NoStop}%
\bibitem [{\citenamefont {Gallo}\ \emph {et~al.}(2012)\citenamefont {Gallo},
  \citenamefont {Rovere},\ and\ \citenamefont {Chen}}]{Gallo:2012}%
  \BibitemOpen
  \bibfield  {author} {\bibinfo {author} {\bibfnamefont {P.}~\bibnamefont
  {Gallo}}, \bibinfo {author} {\bibfnamefont {M.}~\bibnamefont {Rovere}}, \
  and\ \bibinfo {author} {\bibfnamefont {S.-H.}\ \bibnamefont {Chen}},\ }\href
  {http://stacks.iop.org/0953-8984/24/i=6/a=064109} {\bibfield  {journal}
  {\bibinfo  {journal} {J. Phys.: Condens. Matter}\ }\textbf {\bibinfo {volume}
  {24}},\ \bibinfo {pages} {064109} (\bibinfo {year} {2012})}\BibitemShut
  {NoStop}%
\bibitem [{\citenamefont {Krishnan}\ and\ \citenamefont
  {Ayappa}(2012)}]{Krishnan:2012}%
  \BibitemOpen
  \bibfield  {author} {\bibinfo {author} {\bibfnamefont {S.~H.}\ \bibnamefont
  {Krishnan}}\ and\ \bibinfo {author} {\bibfnamefont {K.~G.}\ \bibnamefont
  {Ayappa}},\ }\href {\doibase 10.1103/PhysRevE.86.011504} {\bibfield
  {journal} {\bibinfo  {journal} {Phys. Rev. E}\ }\textbf {\bibinfo {volume}
  {86}},\ \bibinfo {pages} {011504} (\bibinfo {year} {2012})}\BibitemShut
  {NoStop}%
\bibitem [{\citenamefont {Mittal}\ \emph {et~al.}(2008)\citenamefont {Mittal},
  \citenamefont {Truskett}, \citenamefont {Errington},\ and\ \citenamefont
  {Hummer}}]{Mittal:2008}%
  \BibitemOpen
  \bibfield  {author} {\bibinfo {author} {\bibfnamefont {J.}~\bibnamefont
  {Mittal}}, \bibinfo {author} {\bibfnamefont {T.~M.}\ \bibnamefont
  {Truskett}}, \bibinfo {author} {\bibfnamefont {J.~R.}\ \bibnamefont
  {Errington}}, \ and\ \bibinfo {author} {\bibfnamefont {G.}~\bibnamefont
  {Hummer}},\ }\href {\doibase 10.1103/PhysRevLett.100.145901} {\bibfield
  {journal} {\bibinfo  {journal} {Phys. Rev. Lett.}\ }\textbf {\bibinfo
  {volume} {100}},\ \bibinfo {pages} {145901} (\bibinfo {year}
  {2008})}\BibitemShut {NoStop}%
\bibitem [{\citenamefont {Mittal}\ \emph {et~al.}(2006)\citenamefont {Mittal},
  \citenamefont {Errington},\ and\ \citenamefont {Truskett}}]{Mittal:2006}%
  \BibitemOpen
  \bibfield  {author} {\bibinfo {author} {\bibfnamefont {J.}~\bibnamefont
  {Mittal}}, \bibinfo {author} {\bibfnamefont {J.~R.}\ \bibnamefont
  {Errington}}, \ and\ \bibinfo {author} {\bibfnamefont {T.~M.}\ \bibnamefont
  {Truskett}},\ }\href {\doibase 10.1103/PhysRevLett.96.177804} {\bibfield
  {journal} {\bibinfo  {journal} {Phys. Rev. Lett.}\ }\textbf {\bibinfo
  {volume} {96}},\ \bibinfo {pages} {177804} (\bibinfo {year}
  {2006})}\BibitemShut {NoStop}%
\bibitem [{\citenamefont {Mittal}\ \emph
  {et~al.}(2007{\natexlab{a}})\citenamefont {Mittal}, \citenamefont
  {Errington},\ and\ \citenamefont {Truskett}}]{Mittal:2007a}%
  \BibitemOpen
  \bibfield  {author} {\bibinfo {author} {\bibfnamefont {J.}~\bibnamefont
  {Mittal}}, \bibinfo {author} {\bibfnamefont {J.~R.}\ \bibnamefont
  {Errington}}, \ and\ \bibinfo {author} {\bibfnamefont {T.~M.}\ \bibnamefont
  {Truskett}},\ }\href {\doibase 10.1021/jp071369e} {\bibfield  {journal}
  {\bibinfo  {journal} {J. Phys. Chem. B}\ }\textbf {\bibinfo {volume} {111}},\
  \bibinfo {pages} {10054} (\bibinfo {year} {2007}{\natexlab{a}})}\BibitemShut
  {NoStop}%
\bibitem [{\citenamefont {Mittal}\ \emph
  {et~al.}(2007{\natexlab{b}})\citenamefont {Mittal}, \citenamefont
  {Errington},\ and\ \citenamefont {Truskett}}]{Mittal:2007b}%
  \BibitemOpen
  \bibfield  {author} {\bibinfo {author} {\bibfnamefont {J.}~\bibnamefont
  {Mittal}}, \bibinfo {author} {\bibfnamefont {J.~R.}\ \bibnamefont
  {Errington}}, \ and\ \bibinfo {author} {\bibfnamefont {T.~M.}\ \bibnamefont
  {Truskett}},\ }\href {\doibase http://dx.doi.org/10.1063/1.2748045}
  {\bibfield  {journal} {\bibinfo  {journal} {J. Chem. Phys.}\ }\textbf
  {\bibinfo {volume} {126}},\ \bibinfo {eid} {244708} (\bibinfo {year}
  {2007}{\natexlab{b}})}\BibitemShut {NoStop}%
\bibitem [{\citenamefont {Goel}\ \emph {et~al.}(2008)\citenamefont {Goel},
  \citenamefont {Krekelberg}, \citenamefont {Errington},\ and\ \citenamefont
  {Truskett}}]{Goel:2008}%
  \BibitemOpen
  \bibfield  {author} {\bibinfo {author} {\bibfnamefont {G.}~\bibnamefont
  {Goel}}, \bibinfo {author} {\bibfnamefont {W.~P.}\ \bibnamefont
  {Krekelberg}}, \bibinfo {author} {\bibfnamefont {J.~R.}\ \bibnamefont
  {Errington}}, \ and\ \bibinfo {author} {\bibfnamefont {T.~M.}\ \bibnamefont
  {Truskett}},\ }\href {\doibase 10.1103/PhysRevLett.100.106001} {\bibfield
  {journal} {\bibinfo  {journal} {Phys. Rev. Lett.}\ }\textbf {\bibinfo
  {volume} {100}},\ \bibinfo {pages} {106001} (\bibinfo {year}
  {2008})}\BibitemShut {NoStop}%
\bibitem [{\citenamefont {Goel}\ \emph {et~al.}(2009)\citenamefont {Goel},
  \citenamefont {Krekelberg}, \citenamefont {Pond}, \citenamefont {Mittal},
  \citenamefont {Shen}, \citenamefont {Errington},\ and\ \citenamefont
  {Truskett}}]{Goel:2009}%
  \BibitemOpen
  \bibfield  {author} {\bibinfo {author} {\bibfnamefont {G.}~\bibnamefont
  {Goel}}, \bibinfo {author} {\bibfnamefont {W.~P.}\ \bibnamefont
  {Krekelberg}}, \bibinfo {author} {\bibfnamefont {M.~J.}\ \bibnamefont
  {Pond}}, \bibinfo {author} {\bibfnamefont {J.}~\bibnamefont {Mittal}},
  \bibinfo {author} {\bibfnamefont {V.~K.}\ \bibnamefont {Shen}}, \bibinfo
  {author} {\bibfnamefont {J.~R.}\ \bibnamefont {Errington}}, \ and\ \bibinfo
  {author} {\bibfnamefont {T.~M.}\ \bibnamefont {Truskett}},\ }\href
  {http://stacks.iop.org/1742-5468/2009/i=04/a=P04006} {\bibfield  {journal}
  {\bibinfo  {journal} {J. Stat. Mech.}\ }\textbf {\bibinfo {volume} {2009}},\
  \bibinfo {pages} {P04006} (\bibinfo {year} {2009})}\BibitemShut {NoStop}%
\bibitem [{\citenamefont {Krekelberg}\ \emph {et~al.}(2013)\citenamefont
  {Krekelberg}, \citenamefont {Siderius}, \citenamefont {Shen}, \citenamefont
  {Truskett},\ and\ \citenamefont {Errington}}]{Krekelberg:2013}%
  \BibitemOpen
  \bibfield  {author} {\bibinfo {author} {\bibfnamefont {W.~P.}\ \bibnamefont
  {Krekelberg}}, \bibinfo {author} {\bibfnamefont {D.~W.}\ \bibnamefont
  {Siderius}}, \bibinfo {author} {\bibfnamefont {V.~K.}\ \bibnamefont {Shen}},
  \bibinfo {author} {\bibfnamefont {T.~M.}\ \bibnamefont {Truskett}}, \ and\
  \bibinfo {author} {\bibfnamefont {J.~R.}\ \bibnamefont {Errington}},\ }\href
  {\doibase 10.1021/la4037327} {\bibfield  {journal} {\bibinfo  {journal}
  {Langmuir}\ }\textbf {\bibinfo {volume} {29}},\ \bibinfo {pages} {14527}
  (\bibinfo {year} {2013})}\BibitemShut {NoStop}%
\bibitem [{\citenamefont {Ingebrigtsen}\ \emph {et~al.}(2013)\citenamefont
  {Ingebrigtsen}, \citenamefont {Errington}, \citenamefont {Truskett},\ and\
  \citenamefont {Dyre}}]{Ingebrigtsen:2013}%
  \BibitemOpen
  \bibfield  {author} {\bibinfo {author} {\bibfnamefont {T.~S.}\ \bibnamefont
  {Ingebrigtsen}}, \bibinfo {author} {\bibfnamefont {J.~R.}\ \bibnamefont
  {Errington}}, \bibinfo {author} {\bibfnamefont {T.~M.}\ \bibnamefont
  {Truskett}}, \ and\ \bibinfo {author} {\bibfnamefont {J.~C.}\ \bibnamefont
  {Dyre}},\ }\href {\doibase 10.1103/PhysRevLett.111.235901} {\bibfield
  {journal} {\bibinfo  {journal} {Phys. Rev. Lett.}\ }\textbf {\bibinfo
  {volume} {111}},\ \bibinfo {pages} {235901} (\bibinfo {year}
  {2013})}\BibitemShut {NoStop}%
\bibitem [{\citenamefont {Voigtmann}\ \emph {et~al.}(2004)\citenamefont
  {Voigtmann}, \citenamefont {Puertas},\ and\ \citenamefont
  {Fuchs}}]{Voigtmann:2004}%
  \BibitemOpen
  \bibfield  {author} {\bibinfo {author} {\bibfnamefont {{\relax
  Th}.}~\bibnamefont {Voigtmann}}, \bibinfo {author} {\bibfnamefont {A.~M.}\
  \bibnamefont {Puertas}}, \ and\ \bibinfo {author} {\bibfnamefont
  {M.}~\bibnamefont {Fuchs}},\ }\href {\doibase 10.1103/PhysRevE.70.061506}
  {\bibfield  {journal} {\bibinfo  {journal} {Phys. Rev. E}\ }\textbf {\bibinfo
  {volume} {70}},\ \bibinfo {pages} {061506} (\bibinfo {year}
  {2004})}\BibitemShut {NoStop}%
\bibitem [{\citenamefont {Flenner}\ and\ \citenamefont
  {Szamel}(2005)}]{Flenner:2005}%
  \BibitemOpen
  \bibfield  {author} {\bibinfo {author} {\bibfnamefont {E.}~\bibnamefont
  {Flenner}}\ and\ \bibinfo {author} {\bibfnamefont {G.}~\bibnamefont
  {Szamel}},\ }\href {\doibase 10.1103/PhysRevE.72.031508} {\bibfield
  {journal} {\bibinfo  {journal} {Phys. Rev. E}\ }\textbf {\bibinfo {volume}
  {72}},\ \bibinfo {pages} {031508} (\bibinfo {year} {2005})}\BibitemShut
  {NoStop}%
\bibitem [{\citenamefont {Voigtmann}(2011)}]{Voigtmann:2011}%
  \BibitemOpen
  \bibfield  {author} {\bibinfo {author} {\bibfnamefont {{\relax
  Th}.}~\bibnamefont {Voigtmann}},\ }\href
  {http://stacks.iop.org/0295-5075/96/i=3/a=36006} {\bibfield  {journal}
  {\bibinfo  {journal} {EPL}\ }\textbf {\bibinfo {volume} {96}},\ \bibinfo
  {pages} {36006} (\bibinfo {year} {2011})}\BibitemShut {NoStop}%
\bibitem [{\citenamefont {Hajnal}\ \emph {et~al.}(2011)\citenamefont {Hajnal},
  \citenamefont {Oettel},\ and\ \citenamefont {Schilling}}]{Hajnal:2011}%
  \BibitemOpen
  \bibfield  {author} {\bibinfo {author} {\bibfnamefont {D.}~\bibnamefont
  {Hajnal}}, \bibinfo {author} {\bibfnamefont {M.}~\bibnamefont {Oettel}}, \
  and\ \bibinfo {author} {\bibfnamefont {R.}~\bibnamefont {Schilling}},\ }\href
  {\doibase 10.1016/j.jnoncrysol.2010.06.039} {\bibfield  {journal} {\bibinfo
  {journal} {J. Non-Cryst. Solids}\ }\textbf {\bibinfo {volume} {357}},\
  \bibinfo {pages} {302 } (\bibinfo {year} {2011})}\BibitemShut {NoStop}%
\bibitem [{\citenamefont {Krüger}\ \emph {et~al.}(2011)\citenamefont
  {Krüger}, \citenamefont {Weysser},\ and\ \citenamefont
  {Fuchs}}]{Krueger:2011}%
  \BibitemOpen
  \bibfield  {author} {\bibinfo {author} {\bibfnamefont {M.}~\bibnamefont
  {Kr\"uger}}, \bibinfo {author} {\bibfnamefont {F.}~\bibnamefont {Weysser}}, \
  and\ \bibinfo {author} {\bibfnamefont {M.}~\bibnamefont {Fuchs}},\ }\href
  {\doibase 10.1140/epje/i2011-11088-5} {\bibfield  {journal} {\bibinfo
  {journal} {The European Physical Journal E}\ }\textbf {\bibinfo {volume}
  {34}},\ \bibinfo {pages} {1} (\bibinfo {year} {2011})}\BibitemShut {NoStop}%
\bibitem [{\citenamefont {Harrer}\ \emph {et~al.}(2012)\citenamefont {Harrer},
  \citenamefont {Winter}, \citenamefont {Horbach}, \citenamefont {Fuchs},\ and\
  \citenamefont {Voigtmann}}]{Harrer:2012}%
  \BibitemOpen
  \bibfield  {author} {\bibinfo {author} {\bibfnamefont {C.~J.}\ \bibnamefont
  {Harrer}}, \bibinfo {author} {\bibfnamefont {D.}~\bibnamefont {Winter}},
  \bibinfo {author} {\bibfnamefont {J.}~\bibnamefont {Horbach}}, \bibinfo
  {author} {\bibfnamefont {M.}~\bibnamefont {Fuchs}}, \ and\ \bibinfo {author}
  {\bibfnamefont {{\relax Th}.}~\bibnamefont {Voigtmann}},\ }\href
  {http://stacks.iop.org/0953-8984/24/i=46/a=464105} {\bibfield  {journal}
  {\bibinfo  {journal} {J. Phys.: Condens. Matter}\ }\textbf {\bibinfo {volume}
  {24}},\ \bibinfo {pages} {464105} (\bibinfo {year} {2012})}\BibitemShut
  {NoStop}%
\bibitem [{\citenamefont {Sperl}\ \emph {et~al.}(2012)\citenamefont {Sperl},
  \citenamefont {Kranz},\ and\ \citenamefont {Zippelius}}]{Sperl:2012}%
  \BibitemOpen
  \bibfield  {author} {\bibinfo {author} {\bibfnamefont {M.}~\bibnamefont
  {Sperl}}, \bibinfo {author} {\bibfnamefont {W.~T.}\ \bibnamefont {Kranz}}, \
  and\ \bibinfo {author} {\bibfnamefont {A.}~\bibnamefont {Zippelius}},\ }\href
  {http://stacks.iop.org/0295-5075/98/i=2/a=28001} {\bibfield  {journal}
  {\bibinfo  {journal} {EPL}\ }\textbf {\bibinfo {volume} {98}},\ \bibinfo
  {pages} {28001} (\bibinfo {year} {2012})}\BibitemShut {NoStop}%
\bibitem [{\citenamefont {Chong}\ \emph {et~al.}(2001)\citenamefont {Chong},
  \citenamefont {G\"otze},\ and\ \citenamefont {Mayr}}]{Chong:2001}%
  \BibitemOpen
  \bibfield  {author} {\bibinfo {author} {\bibfnamefont {S.-H.}\ \bibnamefont
  {Chong}}, \bibinfo {author} {\bibfnamefont {W.}~\bibnamefont {G\"otze}}, \
  and\ \bibinfo {author} {\bibfnamefont {M.~R.}\ \bibnamefont {Mayr}},\ }\href
  {\doibase 10.1103/PhysRevE.64.011503} {\bibfield  {journal} {\bibinfo
  {journal} {Phys. Rev. E}\ }\textbf {\bibinfo {volume} {64}},\ \bibinfo
  {pages} {011503} (\bibinfo {year} {2001})}\BibitemShut {NoStop}%
\bibitem [{\citenamefont {Chong}\ and\ \citenamefont
  {G\"otze}(2002)}]{Chong:2002}%
  \BibitemOpen
  \bibfield  {author} {\bibinfo {author} {\bibfnamefont {S.-H.}\ \bibnamefont
  {Chong}}\ and\ \bibinfo {author} {\bibfnamefont {W.}~\bibnamefont
  {G\"otze}},\ }\href {\doibase 10.1103/PhysRevE.65.041503} {\bibfield
  {journal} {\bibinfo  {journal} {Phys. Rev. E}\ }\textbf {\bibinfo {volume}
  {65}},\ \bibinfo {pages} {041503} (\bibinfo {year} {2002})}\BibitemShut
  {NoStop}%
\bibitem [{\citenamefont {Krakoviack}(2009)}]{Krakoviack:2009}%
  \BibitemOpen
  \bibfield  {author} {\bibinfo {author} {\bibfnamefont {V.}~\bibnamefont
  {Krakoviack}},\ }\href {\doibase 10.1103/PhysRevE.79.061501} {\bibfield
  {journal} {\bibinfo  {journal} {Phys. Rev. E}\ }\textbf {\bibinfo {volume}
  {79}},\ \bibinfo {eid} {061501} (\bibinfo {year} {2009})}\BibitemShut
  {NoStop}%
\bibitem [{\citenamefont {Spanner}\ \emph {et~al.}(2013)\citenamefont
  {Spanner}, \citenamefont {Schnyder}, \citenamefont {H{\"o}f\/ling},
  \citenamefont {Voigtmann},\ and\ \citenamefont {Franosch}}]{Spanner:2013}%
  \BibitemOpen
  \bibfield  {author} {\bibinfo {author} {\bibfnamefont {M.}~\bibnamefont
  {Spanner}}, \bibinfo {author} {\bibfnamefont {S.~K.}\ \bibnamefont
  {Schnyder}}, \bibinfo {author} {\bibfnamefont {F.}~\bibnamefont
  {H{\"o}f\/ling}}, \bibinfo {author} {\bibfnamefont {{\relax
  Th}.}~\bibnamefont {Voigtmann}}, \ and\ \bibinfo {author} {\bibfnamefont
  {T.}~\bibnamefont {Franosch}},\ }\href {\doibase 10.1039/C2SM27060A}
  {\bibfield  {journal} {\bibinfo  {journal} {Soft Matter}\ }\textbf {\bibinfo
  {volume} {9}},\ \bibinfo {pages} {1604} (\bibinfo {year} {2013})}\BibitemShut
  {NoStop}%
\bibitem [{\citenamefont {Lang}\ \emph {et~al.}(2010)\citenamefont {Lang},
  \citenamefont {Bo\ifmmode~\mbox{\c{t}}\else \c{t}\fi{}an}, \citenamefont
  {Oettel}, \citenamefont {Hajnal}, \citenamefont {Franosch},\ and\
  \citenamefont {Schilling}}]{Lang:2010}%
  \BibitemOpen
  \bibfield  {author} {\bibinfo {author} {\bibfnamefont {S.}~\bibnamefont
  {Lang}}, \bibinfo {author} {\bibfnamefont {V.}~\bibnamefont
  {Bo\ifmmode~\mbox{\c{t}}\else \c{t}\fi{}an}}, \bibinfo {author}
  {\bibfnamefont {M.}~\bibnamefont {Oettel}}, \bibinfo {author} {\bibfnamefont
  {D.}~\bibnamefont {Hajnal}}, \bibinfo {author} {\bibfnamefont
  {T.}~\bibnamefont {Franosch}}, \ and\ \bibinfo {author} {\bibfnamefont
  {R.}~\bibnamefont {Schilling}},\ }\href {\doibase
  10.1103/PhysRevLett.105.125701} {\bibfield  {journal} {\bibinfo  {journal}
  {Phys. Rev. Lett.}\ }\textbf {\bibinfo {volume} {105}},\ \bibinfo {pages}
  {125701} (\bibinfo {year} {2010})}\BibitemShut {NoStop}%
\bibitem [{\citenamefont {Lang}\ \emph {et~al.}(2012)\citenamefont {Lang},
  \citenamefont {Schilling}, \citenamefont {Krakoviack},\ and\ \citenamefont
  {Franosch}}]{Lang:2012}%
  \BibitemOpen
  \bibfield  {author} {\bibinfo {author} {\bibfnamefont {S.}~\bibnamefont
  {Lang}}, \bibinfo {author} {\bibfnamefont {R.}~\bibnamefont {Schilling}},
  \bibinfo {author} {\bibfnamefont {V.}~\bibnamefont {Krakoviack}}, \ and\
  \bibinfo {author} {\bibfnamefont {T.}~\bibnamefont {Franosch}},\ }\href
  {\doibase 10.1103/PhysRevE.86.021502} {\bibfield  {journal} {\bibinfo
  {journal} {Phys. Rev. E}\ }\textbf {\bibinfo {volume} {86}},\ \bibinfo
  {pages} {021502} (\bibinfo {year} {2012})}\BibitemShut {NoStop}%
\bibitem [{\citenamefont {Biroli}\ \emph {et~al.}(2006)\citenamefont {Biroli},
  \citenamefont {Bouchaud}, \citenamefont {Miyazaki},\ and\ \citenamefont
  {Reichman}}]{Biroli:2006}%
  \BibitemOpen
  \bibfield  {author} {\bibinfo {author} {\bibfnamefont {G.}~\bibnamefont
  {Biroli}}, \bibinfo {author} {\bibfnamefont {J.-P.}\ \bibnamefont
  {Bouchaud}}, \bibinfo {author} {\bibfnamefont {K.}~\bibnamefont {Miyazaki}},
  \ and\ \bibinfo {author} {\bibfnamefont {D.~R.}\ \bibnamefont {Reichman}},\
  }\href {\doibase 10.1103/PhysRevLett.97.195701} {\bibfield  {journal}
  {\bibinfo  {journal} {Phys. Rev. Lett.}\ }\textbf {\bibinfo {volume} {97}},\
  \bibinfo {pages} {195701} (\bibinfo {year} {2006})}\BibitemShut {NoStop}%
\bibitem [{\citenamefont {Nandi}\ \emph {et~al.}(2011)\citenamefont {Nandi},
  \citenamefont {Bhattacharyya},\ and\ \citenamefont {Ramaswamy}}]{Nandi:2011}%
  \BibitemOpen
  \bibfield  {author} {\bibinfo {author} {\bibfnamefont {S.~K.}\ \bibnamefont
  {Nandi}}, \bibinfo {author} {\bibfnamefont {S.~M.}\ \bibnamefont
  {Bhattacharyya}}, \ and\ \bibinfo {author} {\bibfnamefont {S.}~\bibnamefont
  {Ramaswamy}},\ }\href {\doibase 10.1103/PhysRevE.84.061501} {\bibfield
  {journal} {\bibinfo  {journal} {Phys. Rev. E}\ }\textbf {\bibinfo {volume}
  {84}},\ \bibinfo {pages} {061501} (\bibinfo {year} {2011})}\BibitemShut
  {NoStop}%
\bibitem [{\citenamefont {Hansen}\ and\ \citenamefont
  {McDonald}(2006)}]{Hansen:Theory_of_Simple_Liquids}%
  \BibitemOpen
  \bibfield  {author} {\bibinfo {author} {\bibfnamefont {J.~P.}\ \bibnamefont
  {Hansen}}\ and\ \bibinfo {author} {\bibfnamefont {I.~R.}\ \bibnamefont
  {McDonald}},\ }\href@noop {} {\emph {\bibinfo {title} {Theory of Simple
  Liquids}}}\ (\bibinfo  {publisher} {Academic Press},\ \bibinfo {year}
  {2006})\BibitemShut {NoStop}%
\bibitem [{\citenamefont {Lang}\ \emph {et~al.}(2013)\citenamefont {Lang},
  \citenamefont {Schilling},\ and\ \citenamefont {Franosch}}]{Lang:2013}%
  \BibitemOpen
  \bibfield  {author} {\bibinfo {author} {\bibfnamefont {S.}~\bibnamefont
  {Lang}}, \bibinfo {author} {\bibfnamefont {R.}~\bibnamefont {Schilling}}, \
  and\ \bibinfo {author} {\bibfnamefont {T.}~\bibnamefont {Franosch}},\ }\href
  {http://stacks.iop.org/1742-5468/2013/i=12/a=P12007} {\bibfield  {journal}
  {\bibinfo  {journal} {J. Stat. Mech.}\ }\textbf {\bibinfo {volume} {2013}},\
  \bibinfo {pages} {P12007} (\bibinfo {year} {2013})}\BibitemShut {NoStop}%
\bibitem [{\citenamefont {Schilling}\ and\ \citenamefont
  {Scheidsteger}(1997)}]{Scheidsteger:1997}%
  \BibitemOpen
  \bibfield  {author} {\bibinfo {author} {\bibfnamefont {R.}~\bibnamefont
  {Schilling}}\ and\ \bibinfo {author} {\bibfnamefont {T.}~\bibnamefont
  {Scheidsteger}},\ }\href {\doibase 10.1103/PhysRevE.56.2932} {\bibfield
  {journal} {\bibinfo  {journal} {Phys. Rev. E}\ }\textbf {\bibinfo {volume}
  {56}},\ \bibinfo {pages} {2932} (\bibinfo {year} {1997})}\BibitemShut
  {NoStop}%
\bibitem [{\citenamefont {Franosch}\ \emph {et~al.}(2012)\citenamefont
  {Franosch}, \citenamefont {Lang},\ and\ \citenamefont
  {Schilling}}]{Franosch:2012}%
  \BibitemOpen
  \bibfield  {author} {\bibinfo {author} {\bibfnamefont {T.}~\bibnamefont
  {Franosch}}, \bibinfo {author} {\bibfnamefont {S.}~\bibnamefont {Lang}}, \
  and\ \bibinfo {author} {\bibfnamefont {R.}~\bibnamefont {Schilling}},\ }\href
  {\doibase 10.1103/PhysRevLett.109.240601} {\bibfield  {journal} {\bibinfo
  {journal} {Phys. Rev. Lett.}\ }\textbf {\bibinfo {volume} {109}},\ \bibinfo
  {pages} {240601} (\bibinfo {year} {2012})}\BibitemShut {NoStop}%
\bibitem [{\citenamefont {Lang}\ \emph {et~al.}(2014)\citenamefont {Lang},
  \citenamefont {Franosch},\ and\ \citenamefont {Schilling}}]{Lang:2014}%
  \BibitemOpen
  \bibfield  {author} {\bibinfo {author} {\bibfnamefont {S.}~\bibnamefont
  {Lang}}, \bibinfo {author} {\bibfnamefont {T.}~\bibnamefont {Franosch}}, \
  and\ \bibinfo {author} {\bibfnamefont {R.}~\bibnamefont {Schilling}},\ }\href
  {\doibase http://dx.doi.org/10.1063/1.4867284} {\bibfield  {journal}
  {\bibinfo  {journal} {J. Chem. Phys.}\ }\textbf {\bibinfo {volume} {140}},\
  \bibinfo {eid} {104506} (\bibinfo {year} {2014})}\BibitemShut {NoStop}%
\bibitem [{\citenamefont {Bayer}\ \emph {et~al.}(2007)\citenamefont {Bayer},
  \citenamefont {Brader}, \citenamefont {Ebert}, \citenamefont {Fuchs},
  \citenamefont {Lange}, \citenamefont {Maret}, \citenamefont {Schilling},
  \citenamefont {Sperl},\ and\ \citenamefont {Wittmer}}]{Bayer:2007}%
  \BibitemOpen
  \bibfield  {author} {\bibinfo {author} {\bibfnamefont {M.}~\bibnamefont
  {Bayer}}, \bibinfo {author} {\bibfnamefont {J.~M.}\ \bibnamefont {Brader}},
  \bibinfo {author} {\bibfnamefont {F.}~\bibnamefont {Ebert}}, \bibinfo
  {author} {\bibfnamefont {M.}~\bibnamefont {Fuchs}}, \bibinfo {author}
  {\bibfnamefont {E.}~\bibnamefont {Lange}}, \bibinfo {author} {\bibfnamefont
  {G.}~\bibnamefont {Maret}}, \bibinfo {author} {\bibfnamefont
  {R.}~\bibnamefont {Schilling}}, \bibinfo {author} {\bibfnamefont
  {M.}~\bibnamefont {Sperl}}, \ and\ \bibinfo {author} {\bibfnamefont {J.~P.}\
  \bibnamefont {Wittmer}},\ }\href {\doibase 10.1103/PhysRevE.76.011508}
  {\bibfield  {journal} {\bibinfo  {journal} {Phys. Rev. E}\ }\textbf {\bibinfo
  {volume} {76}},\ \bibinfo {pages} {011508} (\bibinfo {year}
  {2007})}\BibitemShut {NoStop}%
\bibitem [{\citenamefont {Hajnal}\ \emph {et~al.}(2009)\citenamefont {Hajnal},
  \citenamefont {Brader},\ and\ \citenamefont {Schilling}}]{Hajnal:2009}%
  \BibitemOpen
  \bibfield  {author} {\bibinfo {author} {\bibfnamefont {D.}~\bibnamefont
  {Hajnal}}, \bibinfo {author} {\bibfnamefont {J.~M.}\ \bibnamefont {Brader}},
  \ and\ \bibinfo {author} {\bibfnamefont {R.}~\bibnamefont {Schilling}},\
  }\href {\doibase 10.1103/PhysRevE.80.021503} {\bibfield  {journal} {\bibinfo
  {journal} {Phys. Rev. E}\ }\textbf {\bibinfo {volume} {80}},\ \bibinfo
  {pages} {021503} (\bibinfo {year} {2009})}\BibitemShut {NoStop}%
\bibitem [{\citenamefont {Franosch}\ and\ \citenamefont
  {G{\"o}tze}(1994)}]{Franosch:1994}%
  \BibitemOpen
  \bibfield  {author} {\bibinfo {author} {\bibfnamefont {T.}~\bibnamefont
  {Franosch}}\ and\ \bibinfo {author} {\bibfnamefont {W.}~\bibnamefont
  {G{\"o}tze}},\ }\href {http://stacks.iop.org/0953-8984/6/i=26/a=004}
  {\bibfield  {journal} {\bibinfo  {journal} {J. Phys.: Condens. Matter}\
  }\textbf {\bibinfo {volume} {6}},\ \bibinfo {pages} {4807} (\bibinfo {year}
  {1994})}\BibitemShut {NoStop}%
\bibitem [{\citenamefont {Franosch}\ and\ \citenamefont
  {Singh}(1997)}]{Franosch:1997c}%
  \BibitemOpen
  \bibfield  {author} {\bibinfo {author} {\bibfnamefont {T.}~\bibnamefont
  {Franosch}}\ and\ \bibinfo {author} {\bibfnamefont {A.~P.}\ \bibnamefont
  {Singh}},\ }\href@noop {} {\bibfield  {journal} {\bibinfo  {journal} {J.
  Chem. Phys.}\ }\textbf {\bibinfo {volume} {107}}(\bibinfo {year}  {1997})}\BibitemShut {NoStop}%
\bibitem [{\citenamefont {Franosch}\ and\ \citenamefont
  {Singh}(1999)}]{Franosch:1999b}%
  \BibitemOpen
  \bibfield  {author} {\bibinfo {author} {\bibfnamefont {T.}~\bibnamefont
  {Franosch}}\ and\ \bibinfo {author} {\bibfnamefont {A.~P.}\ \bibnamefont
  {Singh}},\ }\href@noop {} {\bibfield  {journal} {\bibinfo  {journal} {J.
  Chem. Phys.}\ }\textbf {\bibinfo {volume} {110}} (\bibinfo {year}
  {1999})}\BibitemShut {NoStop}%
\end{thebibliography}
%
\end{document}